\documentclass[a4paper,10pt]{article}
\usepackage[pdftex]{graphicx,hyperref}
\usepackage{amsmath,amsfonts,amssymb}
\usepackage{fullpage}
\usepackage{authblk}
\usepackage{setspace}
\usepackage{subcaption}
\usepackage{booktabs}
\usepackage{tabularx}
\usepackage{makecell}
\usepackage[explicit]{titlesec}
\usepackage{sidecap}
\usepackage{pbox}
\usepackage{cite}
\usepackage{amsfonts}
\usepackage{amsmath}
\usepackage{multirow}
\usepackage{longtable}
\usepackage{color}

\newcommand{\beginsupplement}{%
        \setcounter{table}{0}
        \renewcommand{\thetable}{S\arabic{table}}%
        \setcounter{figure}{0}
        \renewcommand{\thefigure}{S\arabic{figure}}%
     }

\title{\bf{Understanding the interplay \\ between social and spatial behaviour}}

\author[a,*]{Laura Alessandretti}
\author[a,b]{Sune Lehmann}
\author[c]{Andrea Baronchelli}

\affil[a]{Technical University of Denmark, DK-2800 Kgs. Lyngby, Denmark}
\affil[b]{{Niels Bohr Institute, University of Copenhagen, DK-2100 K\o benhavn \O, Denmark }
\affil[c]{City,  University of London, London EC1V 0HB, United Kingdom}
\newline
{*\small{l.alessandretti@gmail.com}}}
\date{}
\begin{document}
\maketitle

\begin{abstract} 
According to personality psychology, personality traits determine many aspects of human behaviour. However, validating this insight in large groups has been challenging so far, due to the scarcity of multi-channel data. 
Here, we focus on the relationship between mobility and social behaviour by analysing trajectories and mobile phone interactions of $\sim 1,000$ individuals from two high-resolution longitudinal datasets. We identify a connection between the way in which individuals explore new resources and exploit known assets in the social and spatial spheres. We show that different individuals balance the exploration-exploitation trade-off in different ways and we explain part of the variability in the data by the big five personality traits. We point out that, in both realms, extraversion correlates with the attitude towards exploration and routine diversity, while neuroticism and openness account for the tendency to evolve routine over long time-scales. We find no evidence for the existence of classes of individuals across the spatio-social domains. Our results bridge the fields of human geography, sociology and personality psychology and can help improve current models of mobility and tie formation.
\end{abstract}

\section*{Introduction}

Our social and spatial behaviour are shaped by both internal and external constraints. On one hand,  external factors \cite{hagerstraand1970people} such as time, cognition, age or the need for food constrain our possibilities. On the other hand, we are driven by internal needs, purposes and preferences. Specifically, within personality psychology, it has been conjectured that personality traits play a key role in shaping our choices across various situations \cite{mischel1973toward,allport1937personality}.

In the social realm, individuals cope with cognitive and temporal constraints by establishing and maintaining connections in a distinctive \cite{saramaki2014persistence,miritello2013limited} and persistent \cite{saramaki2014persistence} manner. For example, the size of an individual's social circle is bounded under $\sim 150$, the so-called Dunbar number \cite{dunbar1993coevolution}, but varies among individuals around this limit \cite{roberts2009exploring}. These differences result from an interplay between physical and extrinsic factors such as gender \cite{igarashi2005gender}, age \cite{wrzus2013social} and socio-economic status \cite{mcpherson2006social} as well as from stable individual dispositions underlying personality \cite{roberts2008individual}.

Spatially, individuals are characterised by an \emph{activity space} of repeatedly visited locations within which they move during their daily activities \cite{perchoux2013conceptualization}, but this geo-spatial signature varies in size \cite{gonzalez2008understanding} and spatial shape \cite{pappalardo2015returners}. However, unlike the social case, the conjecture that individuals' spatial behaviour is persistent in time \cite{aitken1991person} had not been verified until recently. 

Here, we capitalise on the recent discovery that the size of the activity space is conserved and correlates with the social circle size \cite{alessandretti2016evidence} to test the conjecture that the same personality dispositions in part determine social and spatial behaviour. We test this theory by analysing two long-term datasets consisting of $\sim 1000$ individuals mobility trajectories and their phone interactions (for previous studies see section `\nameref{State of the art}' below). 

First, we test the hypothesis that the strategies individuals adopt in order to choose where to go and with whom to interact are similar. Then, we identify and characterise the prevailing socio-spatial profiles appearing in the datasets. Finally, we show that socio-spatial profiles can be partially explained by the widely adopted big-five personality trait model, often used to describe aspects of the social and emotional life \cite{lu2009size,casciaro1998seeing,roberts2008individual,roberts2009exploring,hojat1982loneliness,branje2004relations,amato1990personality,hamburger2000relationship}. In the section `\nameref{State of the art}', we review the relevant literature; in `\nameref{Methods}' we describe data collection and pre-processing, and we provide details of the methods implemented; in `\nameref{Results}' we present our findings.

\section*{State of the art}
\label{State of the art}

Individual-level variability in social and spatial behaviour has mostly been investigated in isolation so far, with few notable efforts to reconcile the two. Here, we briefly review the empirical findings in the two domains.

\subsection*{The social domain}
Individuals deal with limited time and cognitive capacity resulting in finite social networks \cite{dunbar1993coevolution,gonccalves2011modeling} by distributing time unevenly across their social circle \cite{sutcliffe2012relationships,arnaboldi2012analysis, miritello2013time,
saramaki2014persistence,zhou2005discrete,marlow2009maintained}. While this is a shared strategy, there is clear evidence for individual-level variation. 
First, social circles vary in terms of diversity: they differ in size \cite{roberts2009exploring} - within a maximum upper-bound of $\sim 150$ individuals \cite{dunbar1993coevolution} -  and in structure \cite{saramaki2014persistence,centellegher2017personality}. Second, individuals display different attitudes towards exploration of social opportunities as they are more or less keen on creating new connections \cite{wehrli2008personality,kumar2010structure,newman2001clustering,mislove2008growth}.  Finally, individuals manage social interactions over time in different ways. Some are characterised by high level of stability as they maintain a very stable social circle, while others renew their social ties at high pace \cite{miritello2013limited}. 

These heterogeneities can be partially explained by factors including gender \cite{dunbar1995social,igarashi2005gender}, age \cite{wrzus2013social, carrasco2008far,van2009size}, socio-economic status \cite{campbell1986social,mcpherson2006social} and physical 
attractiveness \cite{reis1980physical}. Moreover, as conjectured by personality psychologists \cite{jaccard1974predicting,mischel1973toward}, differences in personalities partially explain the variability in social circle composition \cite{roberts2008individual,roberts2009exploring, staiano2012friends,lu2009size,asendorpf2003personality,kalish2006psychological,pollet2011extraverts,de2013predicting}, and the different attitudes towards forming \cite{wehrli2008personality,selfhout2010emerging}, developing \cite{asendorpf2006predictive,branje2004relations} and replacing \cite{centellegher2017personality} social connections. It is worth noticing that many of these findings are recent, resulting from the analysis of digital communication traces.

\subsection*{The spatial domain}
Constraints including physical capabilities, the distribution of resources, and the need to coordinate with others limit our possibilities to move in space \cite{hagerstraand1970people}. Individuals cope with these limitations by allocating their time within an activity space of repeatedly visited locations \cite{golledge1997spatial}, whose size is conserved over several years according to a recent study based on high-resolution trajectories \cite{alessandretti2016evidence}, and previous ones based on unevenly sampled and low spatial resolution data  \cite{jarv2014understanding,schonfelder2010urban}. The activity space varies across individuals in terms of size \cite{alessandretti2016evidence} and shape \cite{pappalardo2015returners}: it was shown that two distinct classes of individuals, \emph{returners} and \emph{explorers}, can be identified based on their propensity to visit new locations, similarly to the social domain \cite{miritello2013limited}. Heterogeneities in spatial behaviour can be explained in terms of gender \cite{kwan2000gender}, age \cite{vazquez2013using,kang2010analyzing},  socio-economic \cite{zenk2011activity, carrasco2008far} and ethnic \cite{kwan2004geovisualization} differences. There has only been sporadic efforts to include personality measures in geographic research, despite the strong connections between the two \cite{van2010transport}. Recent works \cite{chorley2015personality, de2013predicting} suggest that spatial behaviour can be partially explained from personality traits. However, in \cite{chorley2015personality}, this understanding is based on biased data collected from location-based social networks. In \cite{de2013predicting}, the connection between spatial behaviour and personality is not investigated extensively, as it is not the main focus of the study.

\subsection*{Social and spatial connection}
Recently, connections between the social and spatial behaviour of pairs \cite{backstrom2010find,mcgee2013location,sadilek2012finding,crandall2010inferring,
grabowicz2014entangling,toole2015coupling} and groups \cite{onnela2011geographic} of individuals have been demonstrated, and used to design predictive models of mobility \cite{mcgee2013location,jurgens2013s,cho2011friendship} or social ties \cite{wang2011human,sadilek2012finding,scellato2011exploiting,pham2013ebm}. 
Shifting the attention to the individual level, recent works based on online social network data  \cite{cheng2011exploring,cranshaw2010bridging}, mobile phone calls data \cite{toole2015coupling} and evenly sampled high resolution mobility trajectories \cite{alessandretti2016evidence} have shown correlations between the activity space size and the ego network structure, calling for further research to more closely examine the connections between social and spatial behaviour at the individual level.

\section*{Methods}
\label{Methods}

\subsection*{Data description and pre-processing}
Our study is based on $850$ high resolution trajectories and call records of participants in a  $24$ months longitudinal experiment, the Copenhagen Networks Study (CNS) \cite{stopczynski2014measuring}. Results on the connections between social and spatial behaviour were corroborated with data from another experiment with fixed rate temporal sampling, but lower spatial resolution and sample size: the Lausanne Mobile Data Challenge (MDC)~\cite{kiukkonen2010towards, laurila2012mobile}, lasted for $19$ months (see Table~\ref{table_data}).

\renewcommand*{\arraystretch}{1.5}

\begin{table}
\centering

\begin{tabularx}{\columnwidth}{@{}X rrrrr@{}}
 & N & $\delta t$ & $T$ & $\delta x$ & $TC$\\
\midrule
CNS & 850 & 16 s & 24 months & 10 m & 0.84 \\
MDC & 185 &  60 s & 19 months &   100-200 m &0.73  \\
\bottomrule
\end{tabularx}

\caption{\textbf{Characteristics of the  mobility datasets considered.} $N$ is the number of individuals,  $\delta t$ the temporal resolution, $T$ the duration of data collection, $\delta x$ the spatial resolution, $TC$ the median weekly time coverage, defined as the fraction of time an individual's location is known.}
\label{table_data}
\end{table}

\subsubsection*{CNS dataset} The Copenhagen Networks Study (CNS) experiment took place between September 2013 and September 2015 \cite{stopczynski2014measuring} and involved $\sim 1000$ Technical University of Denmark students ($\sim 22\%$ female, $\sim 78\%$ male) typically aged between 19 and 21 years old. Participants' position over time was estimated combining their smart-phones WiFi and GPS data using the method described in~\cite{sapiezynski2015opportunities, alessandretti2016evidence}. The location estimation error is below 50 meters in 95\% of the cases. Participants' calls and sms activity was also collected as part of the experiment. Individuals' background information were obtained through a 310 questions survey including the Big Five Inventory \cite{john1999big}, which measures how individuals score on five broad domains of human personality traits: openness, conscientiousness, extraversion, agreeableness, neuroticism. The personality questionnaire used in the study is a version of the Big Five Inventory \cite{john1999big}, translated from English into Danish. It contains 44 individual items and each trait is computed as the average of 7-10 items. Data collection was approved by the Danish Data Protection Agency. All participants provided individual informed consent. Mobility patterns of participants in the CNS experiment display statistical properties consistent with previous literature \cite{gonzalez2008understanding}, as shown in \cite{alessandretti2016evidence}. \\

\subsubsection*{MDC dataset} Data was collected by the Lausanne Data Collection Campaign between October 2009 and March 2011. The campaign involved an heterogeneous sample of $\sim185$ volunteers with mixed backgrounds from the Lake Geneva region (Switzerland), who were allocated smart-phones \cite{laurila2012mobile}. In this work we used GSM data, that has the highest temporal sampling. Following Nokia's privacy policy, individuals participating in the study provided informed consent \cite{laurila2012mobile}. The Lausanne Mobile Data Challenge experiment involves 62\% male and 38\% female participants, where the age range 22-33 year-old accounts for roughly 2/3 of the population \cite{laurila2013big}.

\subsection*{Metrics}
In this section, we define the concepts and metrics used to quantify the social and spatial behaviour of an individual $i$.

\noindent \textbf{Exploration behaviour} is characterised by the following quantities:
\begin{description}
\item{\textit{Number of new locations/week:}}  $n_{loc}(i,t)$  is the number of locations discovered by $i$ in the week preceding $t$. 
\item{\textit{Number of new ties/week:}} $n_{tie}(i,t)$  is the number of individuals who had contact with $i$ (by sms or call) for the first time in the week preceding $t$.
\end{description}
Note that locations/ties are considered `new' only if discovered after $20$ weeks from the beginning of data collection.

\noindent \textbf{Exploitation behaviour} can be quantified by considering:
\begin{description}
\item{\textit{Activity space:}} The set $AS(i,t)=\{\ell_1,\ell_2,...,\ell_j,...\ell_C\}$ of locations $\ell_j$ that individual $i$ visited at least twice and where she spent a time $\tau_j$ larger than $200 min$ during a time-window of $T=20$ weeks preceding time $t$ (see Supplementary Material for the analysis with $T=30$ weeks). Among the locations in the activity space, $i$ visited $\ell_j$ with probability $p (\ell_j) = \tau_j/\sum \tau_j$. (It is worth noting that this time-based definition of activity space includes all significant locations independently of their spatial position and it is only loosely connected with space-oriented definitions widespread in the geography literature such as the ``standard deviational ellipse" and the ``road network buffer" \cite{sherman2005suite}).

\item{\textit{Social circle: }} The set $SC(i,t)=\{u_1,u_2,...,u_j,...u_k\}$ of individuals $u_j$ with whom individual $i$ had a number of contacts $n_j>5$ by sms or call during a time-window of $T=20$ consecutive weeks preceding time $t$ (see Supplementary Material for the analysis with $T=30$ weeks). The probability that $i$ has contact with a given member  $u_j$ of her social circle is $p(u_j) = n_j/\sum n_j$. 
\end{description}
For these two sets $AS(i,t)$ and $SC(i,t)$, we consider their sizes $C(i,t)$ and $k(i,t)$, quantifying the number of favoured locations and social ties, respectively; their entropies $H_{AS}(i,t)$ and $H_{SC}(i,t)$, measuring how time is allocated among locations and ties; their stabilities $J_{AS}(i,t)$ and $J_{SC}(i,t)$, quantifying the fraction of conserved locations and ties, respectively, across consecutive non-overlapping windows of $T=20$ weeks (see Supplementary Material for $T=30$); their rank turnovers $R_{AS}(i,t)$ and $R_{SC}(i,t)$ measuring the average absolute change in rank of an element in the set between consecutive windows. The mathematical definition of these quantities is provided in Table \ref{table_metrics} 
\begin{table}
\centering

\begin{tabularx}{\columnwidth}{@{}l X l@{}}
  & Activity space & Social circle \\
\midrule
1) Size & $C(i,t) = |AS_i(t)|$ & $k(i,t) = |SC(i,t)|$ \\[3ex]
2) Entropy & $H_{AS} (i,t) = -\sum\limits_{j=1}^{C(i,t)}{p(j) \log p(j)}$ & $H_{SC}(i,t) = -\sum\limits_{j=1}^{k(i,t)}{p(j) \log p(j)}$ \\[4ex]
3) Stability & $J_{SC}(i,t)=\dfrac{|SC(i,t)\cap SC(i,t-T)|}{|SC(i,t)\cup SC(i,t-T)|}^*$ & $J_{AS}(i,t)=\dfrac{|AS(i,t)\cap AS(i,t-T)|}{|AS(i,t)\cup AS(i,t-T)|}^*$ \\[4ex]
\makecell[l]{4) Rank \\
\hspace{7pt} Turnover} & $R_{AS}(i,t)=\sum\limits_{j=1}^{N}\dfrac{|r(j,t) - r(j, t- T)|^{**}}{N}$ & $R_{SC}(i,t)=\sum\limits_{j=1}^{N}\dfrac{|r(j,t) - r(j, t- T)|^{**}}{N}$ \\[4ex]
\midrule
\multicolumn{3}{l}{\small{$^{*}$ Here $T=20$ weeks, see Supplementary Material for the analysis with $T=30$ weeks}} \\
\multicolumn{3}{l}{\small{$^{**}$ $r(\ell_k,t)$ and $r(u_k,t)$ denote the rank of a location $\ell_k$ and individual $u_k$ at $t$, respectively}} \\
\bottomrule
\end{tabularx}
\caption{\textbf{Definition of the metrics characterising the activity space and the social circle.} 1) The size of a set is the number of elements in the set 2) We compute the entropy of a set considering the probability $p(j)$ associated to each element $j$ of the set. 3) We measure the stability  $J_{AS}$ by computing the Jaccard similarity between the activity space at $t$ and at $t-T$, with $T=20$ weeks. $J_{SC}$ is computed in the same way for the social circle. 4) We compute the rank turnover of a set by measuring for each of its elements $j$ the absolute change in rank between two consecutive time windows of length $T=20$ weeks. The rank is attributed based on the probability $p(j)$. The average absolute change in rank across all elements corresponds to the rank turnover. }.
\label{table_metrics}
\end{table}

\subsubsection*{Other metrics} 
In order to compare the difference in entropy between two different sets, we compute their Jensen-Shannon divergence (JSD). The JSD between two sets $P_1$ and $P_2$ is computed as $JSD(P_1,P_2) = H(\frac{1}{2} (P_1+P_2)) - \frac{1}{2} [H(P_1) + H(P_2)]$ (see also \cite{saramaki2014persistence}).

\section*{Results}
\label{Results}
Both in their spatial and social behaviour, individuals are constantly balancing a trade-off between the exploitation of familiar options (such as returning to a favourite restaurant or spending time with an old friend) and the exploration of new opportunities (such as visiting a new bar or going on a first date) \cite{hills2015exploration}. We adopt this exploration-exploitation perspective to analyse the relationship between social and spatial strategies in our dataset \cite{alessandretti2016evidence}.

\begin{figure*}[h!]
\centering
\includegraphics[width=\textwidth]{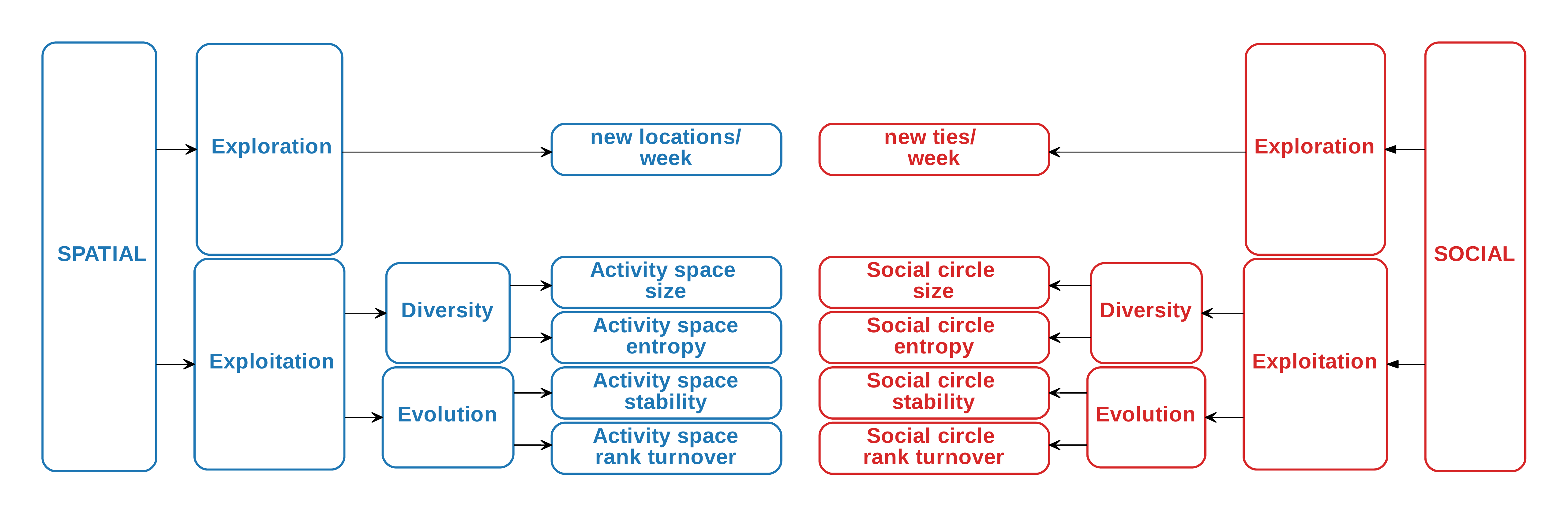}
\caption{\textbf{Schematic description of our framework.}}
\label{schema}
\end{figure*}

We quantify the propensity for exploration and exploitation within each individual, $i$, using the metrics reported in Table \ref{table_metrics_2}, Fig.~\ref{schema} and described in section `\nameref{Methods}'. We focus on two aspects of exploitation, (i) \textit{diversity}, characterising how individuals allocate time among their set of familiar locations and friends, and (ii) \textit{evolution}, characterising the tendency to change exploited locations and friends over time. 

\begin{table}[h!]
\centering
\begin{tabularx}{\columnwidth}{@{}l  X l l @{}}
& {Exploration}  & {Exploitation: Diversity} & {Exploitation: Evolution} \\ 
\midrule 
Spatial &New loc./week, $n_{loc}$ & Activity space size, $C$ & Activity space stability, $J_{AS}$  \\
& & Activity space entropy, $H_{AS}$ & Activity space rank turnover, $R_{AS}$	\\
Social &New ties/week, $n_{tie}$  & Social circle size, $k$ & Social circle stability, $J_{SC}$ \\
& & Social circle entropy, $H_{SC}$ & Social circle rank turnover, $R_{SC}$    \\
\bottomrule
\end{tabularx}
\caption{\textbf{Metrics characterising social and spatial behaviour.} The metrics are defined in section \nameref{Methods}}.
\label{table_metrics_2}
\end{table}

\textbf{Exploration and exploitation are persistent in time.}
First, we verify that individual behaviour is persistent in time. For all the aforementioned measures, we compare the individual self-variation across time $d_{self}(i)$ with a reference difference $d_{ref}(i,j)$ between individuals $i$ and $j$. In the case of the activity space size,  for example, self-variation is measured as $d_{self} = \langle | C(i,t) - C(i,t-T) |\rangle $, where $\langle \cdot \rangle$ is the average across time and $T=20$ weeks (see Supplementary Material for $T=30$) ; the reference difference is computed as $d_{ref}(i,j)= |\langle C(i,t) \rangle - \langle C(j,t) \rangle|$. If $d_{self}(i)<d_{ref}(i,j)$ for most $j$, we can conclude that for individual $i$, fluctuations of the activity space size are negligible compared to the difference with other individuals. The same procedure is followed for all metrics with an adjustment in the case of entropies: The persistence of the entropy $H_{AS}$ is verified by comparing the Janson-Shannon divergences $d_{self} = JSD(AS(i,t), AS(i,t - T))$ and $d_{ref} = JSD(AS(i,t), AS(j,t))$. The same method was used for $H_{SC}$ (see Methods and \cite{saramaki2014persistence}).

Results from the CNS dataset reported in Table \ref{table1_CNS} show that for all metrics $d_{self}(i)<d_{ref}(i,j)$ holds in more than $99\%$ of cases on average (MDC: $97\%$, see Supplementary Material Table S1). Moreover, the average self-variation across the population $\overline{d_{self}}$ is consistent with $\overline{d_{self}}=0$ within errors, and $
\overline{d_{self}}$ significantly smaller than the average reference difference $\overline{d_{ref}}$ (see Tables \ref{table1_CNS} and S1 in Supplementary Material).

\begin{table}[h!]
\centering
\begin{tabularx}{\columnwidth}{@{}X rrr@{}}
{} &          $\overline{d_{self}}$ & $\overline{d_{ref}}$ & $\overline{d_{self}(i)<d_{ref}(i,j)}$ \\
\midrule
Social circle size, $k$                &                $0.04 \pm 0.09$ &           $12 \pm 5$ &                               $99 $\% \\
Activity space size, $C$               &                $0.04 \pm 0.07$ &            $7 \pm 3$ &                               $99 $\% \\
New ties/week, $n_{tie}$          &                $0.05 \pm 0.10$ &        $0.9 \pm 0.5$ &                               $96 $\% \\
New locations/week, $n_{loc}$               &                $0.10 \pm 0.17$ &            $1 \pm 1$ &                               $95 $\% \\
Social circle entropy, $H_{SC}$        &              $0.002 \pm 0.007$ &        $0.7 \pm 0.2$ &                               $99 $\% \\
Activity space entropy, $H_{AS}$       &              $0.002 \pm 0.005$ &        $0.4 \pm 0.1$ &                               $99 $\% \\
Social circle stability, $J_{SC}$      &  $({9} \pm {22}) \cdot10^{-4}$ &      $0.13 \pm 0.05$ &                              $100 $\% \\
Activity space stability, $J_{AS}$     &  $({9} \pm {26}) \cdot10^{-4}$ &      $0.10 \pm 0.04$ &                               $99 $\% \\
Social circle rank turnover, $R_{SC}$  &                $0.05 \pm 0.39$ &            $2 \pm 1$ &                               $99 $\% \\
Activity space rank turnover, $R_{AS}$ &                $0.04 \pm 0.10$ &            $2 \pm 1$ &                               $99 $\% \\
\bottomrule
\end{tabularx}
\caption{\textbf{CNS dataset: Persistence of social and spatial behaviour.}For each of the social and spatial metrics, $\overline{d_{self}}$ is the average self-distance and $\overline{d_{ref}}$ is the reference distance between an individual and all others, averaged across individuals. The third column reports the fraction of cases where $\overline{d_{self}(i)<d_{ref}(i,j)}$, averaged across the population.}
\label{table1_CNS}
\end{table}

These results extend previous findings \cite{saramaki2014persistence, alessandretti2016evidence} and suggest that each individual is characterised by a distinctive socio-spatial behaviour captured by the ensemble of these metrics averaged across time. In fact, these averages are heterogeneously distributed across the samples considered (see Fig.~\ref{distributions}).

\begin{figure*}[h!]
\centering
\includegraphics[width=.9\textwidth]{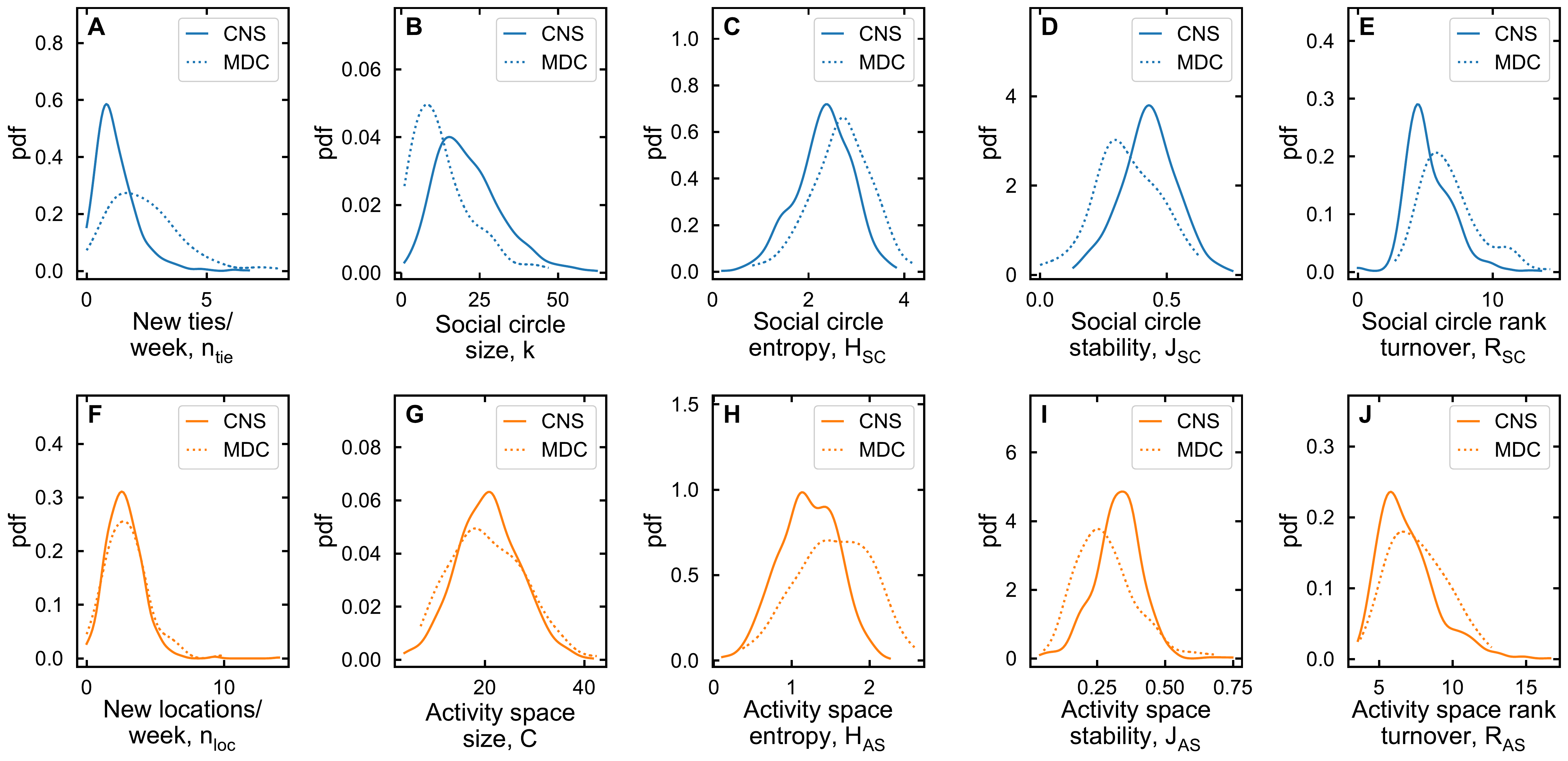}
\caption{\textbf{Distribution of social (above line) and spatial (bottom line) metrics for the CNS and MDC datasets.}}
\label{distributions}
\end{figure*}

\textbf{Exploration and exploitation are correlated in the social and spatial domain.} A natural way to test the interdependency between social and spatial behaviours is measuring the correlation between a given social metric and a corresponding spatial one. We find positive and significant correlations for all metrics and datasets (see Figs.~\ref{social vs spatial_CNS} and S1 in Supplementary Material).

\begin{figure*}[h!]
\centering
\includegraphics[width=.9\textwidth]{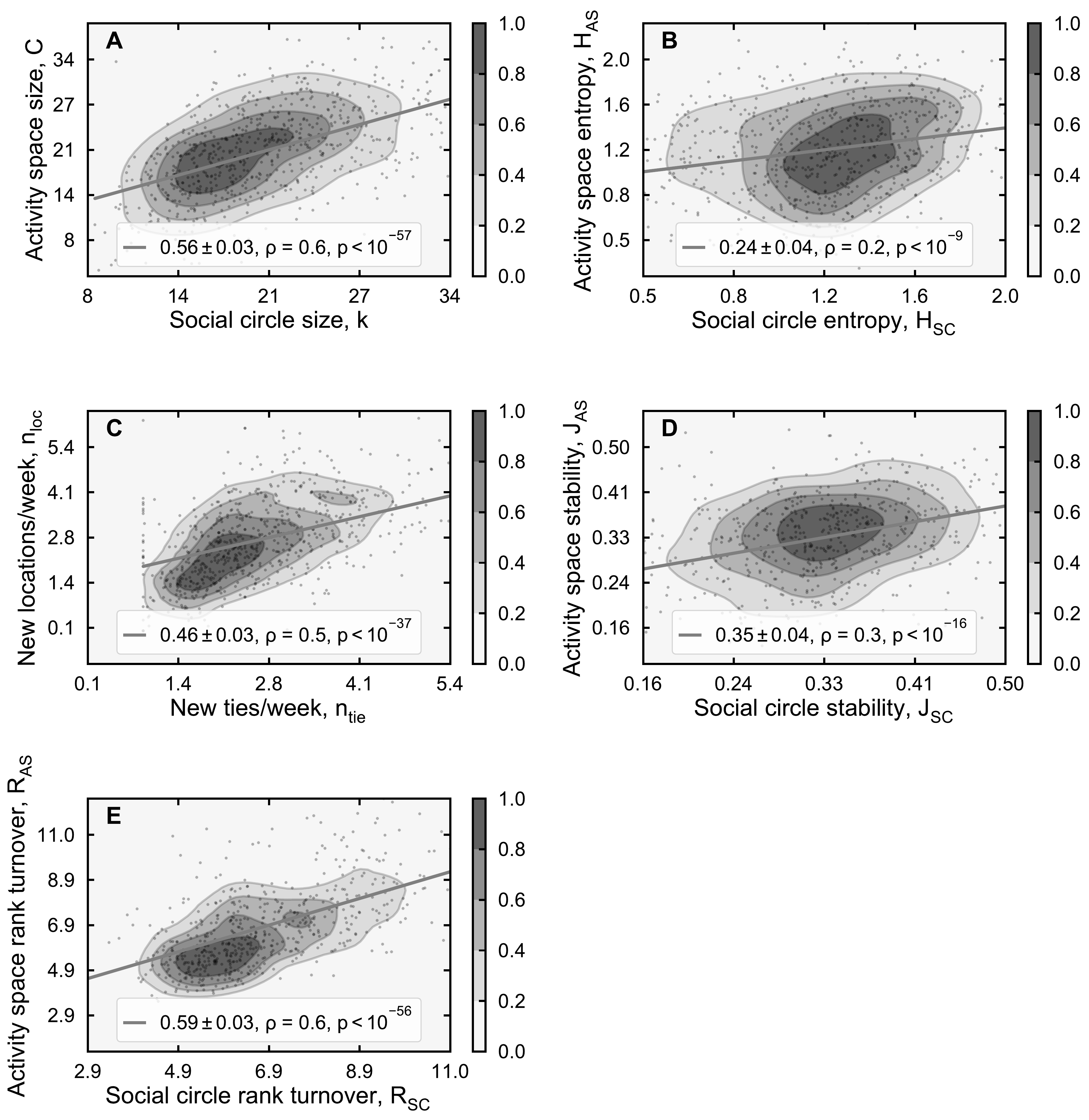}
\caption{\textbf{CNS dataset: correlation between the four dimensions of social and spatial behaviour.} ({A}) Activity space vs social circle size.  ({B}) Activity space vs social circle composition measured as their entropy. ({C}) Average number of new locations vs new ties per week. ({D}) Stability of the activity space vs the stability of the social circle measured as the Jaccard similarity between their composition in consecutive time-windows. ({E}) Rank turnover of the activity space vs the rank turnover of the social circle. Coloured filled areas correspond to cumulative probabilities estimated via Gaussian Kernel Density estimations. Grey lines correspond to linear fit with angular coefficient $b$ reported in the legend. The Pearson correlation coefficient, with corresponding p-value, is reported in the legend. }
\label{social vs spatial_CNS}
\end{figure*}

We find that individuals with high propensity to explore new locations are also more keen on exploring social opportunities (see Fig.~\ref{social vs spatial_CNS}A). Those with diverse mobility routine are also likely to have a correspondingly large social circle (see Fig.~\ref{social vs spatial_CNS}B), and those that often replace social ties, have also an unstable set of favourite locations  (see Fig.~\ref{social vs spatial_CNS}C and D). 

We verify that the observed correlations are not spurious by performing multiple regression analyses that control for other possible sources of variation: gender, age, and time coverage (the average time an individual position is known). We implement five multiple linear regression models M1, M2, M3, M4 and M5. Each regression model predicts a given spatial metric (the activity space size $C$, the activity space entropy $H_{AS}$, the number of new locations/week $n_{loc}$, the activity space stability $J_{AS}$ and the rank turnover $R_{AS}$) using the corresponding social metric and the control variables (age, gender and time coverage) as regressors. The relative importance of each regressor is assessed using the Lindeman, Merenda and Gold ($LMG$) \cite{gromping2006relative} method. 

Results obtained via weighted least square regression (see Tables \ref{multiple regression CNS} for the CNS dataset and S2 in Supplementary Material for the MDC dataset) reveal that the social metrics are significant predictors for spatial metrics (p value$<0.01$ in all cases except for M4 in the MDC dataset), and they typically have more importance than factors such as gender, time, coverage and age group (see Fig.~\ref{multiple regression_figure}). 
\begin{table}
\centering

\begin{tabularx}{\columnwidth}{@{}Xrrr@{}}

\textbf{Model M1:  Activity space size, $C$} &                                      coeff &        p val &   LMG \\
\hline
Social circle size, $k$ &                                          $4 \pm 0$ &  $<10^{-50}$ &  0.94 \\
gender                  &                                     $-0.4 \pm 0.2$ &         0.05 &  0.05 \\
time coverage           &                                      $0.4 \pm 0.2$ &         0.06 &  0.01 \\
{[$R^2=0.32$, $F=100.44$, $p_F=0.0$ ]} 
\\
\\
\textbf{Model M2:  Activity space entropy, $H_{AS}$ } \\
\hline
Social circle entropy, $H_{SC}$ &                                   $0.07 \pm 0.01$ &  $<10^{-6}$ &  0.42 \\
gender                          &                                  $-0.06 \pm 0.01$ &  $<10^{-4}$ &  0.22 \\
time coverage                   &                                  $-0.07 \pm 0.01$ &  $<10^{-5}$ &  0.36 \\
{[$R^2=0.11$, $F=27.30$, $p_F=0.0$]} \\
\\
 \textbf{Model M3:  New locations/week, $n_{loc}$} & 
\\
\hline
New ties/week, $n_{tie}$ &                                   $0.60 \pm 0.05$ &  $<10^{-32}$ &   0.9 \\
gender                        &                                  $-0.16 \pm 0.05$ &   $<10^{-3}$ &  0.08 \\
time coverage                 &                                 $0.001 \pm 0.047$ &          1.0 &  0.01 \\
{[$R^2=0.22$, $F=61.99$, $p_F=0.0$]} &                                              &         &    \\
\\
\textbf{Model M4:  Activity space stability, $J_{AS}$} & \\

\hline
Social circle stability, $J_{SC}$ &                                 $0.024 \pm 0.004$ &  $<10^{-10}$ &   0.6 \\
gender                            &                                 $0.007 \pm 0.003$ &         0.05 &  0.04 \\
time coverage                     &                                 $0.017 \pm 0.004$ &   $<10^{-5}$ &  0.36 \\
{[$R^2=0.16$, $F=33.36$, $p_F=0.0$]} &                                             &         &    \\
\\
 \textbf{Model M5:  Activity space rank turnover, $R_{AS}$} &  \\

\hline
Social circle rank turnover, $R_{SC}$ &                                          $1 \pm 0$ &  $<10^{-56}$ &  0.98 \\
gender                                &                                    $0.12 \pm 0.07$ &         0.06 &  0.01 \\
time coverage                         &                                   $-0.12 \pm 0.07$ &         0.07 &  0.01 \\
{[$R^2=0.36$, $F=108.31$, $p_F=0.0$]} &                                               &        &    \\
\\
\hline
\end{tabularx}
\caption{\textbf{Linear regression models for the CNS dataset.} For each model, we show the coefficients (coeff) calculated by the regression model, the probability (p val) that the variable is not relevant, and the relative importance (LMG) of each regressor computed using the  Lindeman, Merenda and Gold method \cite{gromping2006relative}. Gender is a binary variable taking value 1 for females and 2 for males. For this dataset, age is not relevant as all participants have similar age. For each model, we report the $R^2$ goodness of fit, the $F-test$ statistics with the corresponding p-value $p_F$. }
\label{multiple regression CNS}
\end{table}

\begin{figure}
\includegraphics[width=.9\textwidth]{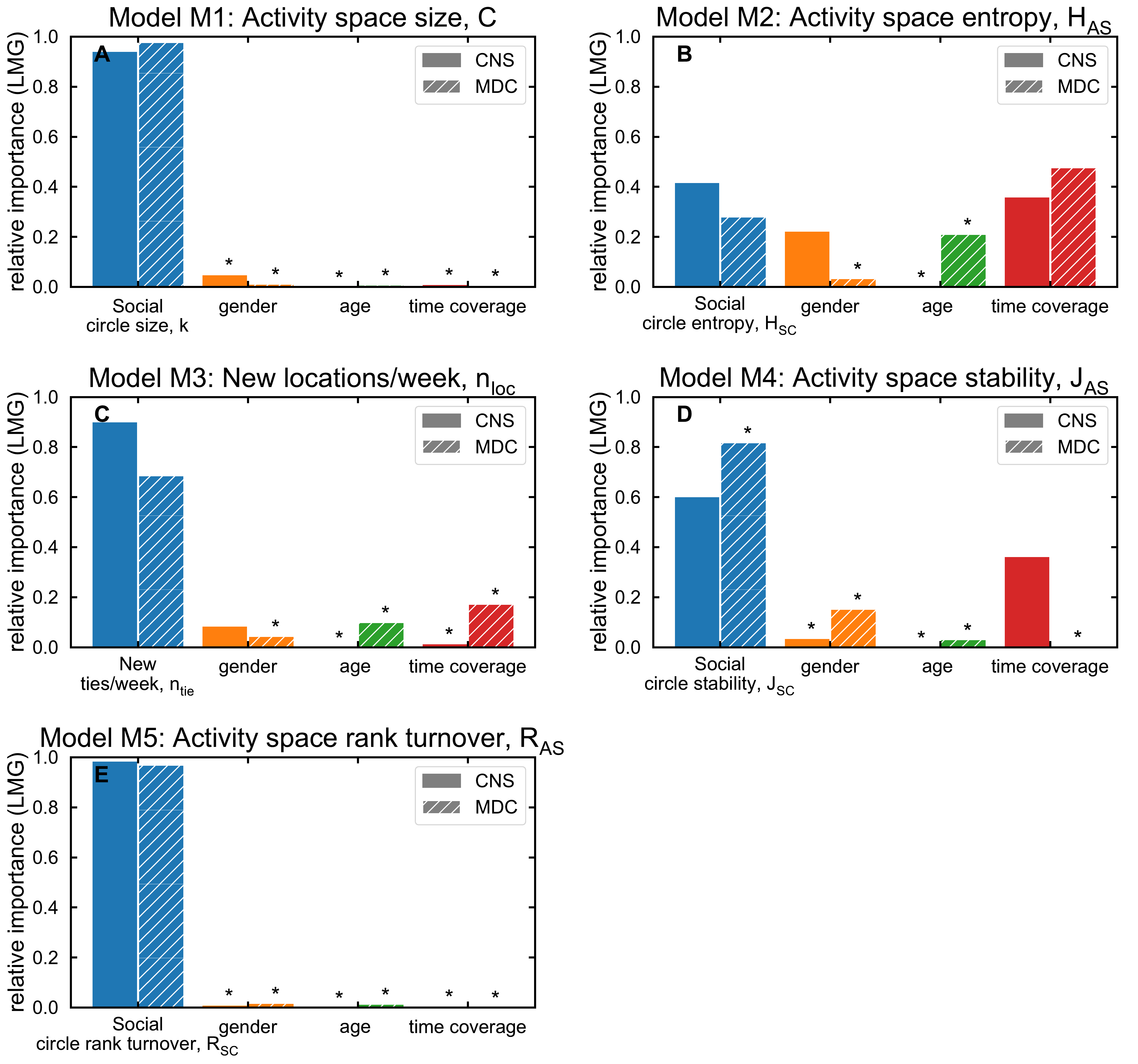}
\caption{\textbf{Relative importance of regressors} LMG of each regressor computed using the  Lindeman, Merenda and Gold method \cite{gromping2006relative} for models M1 (A), M2 (B), M3 (C), M4 (D) and M5 (D). Plain bars show results for the CNS dataset, dashed bars for the MDC dataset. Variables that are not significant in the regression model are marked with *.  }
\label{multiple regression_figure}
\end{figure}

Among the control variables, gender is a significant predictor of spatial behaviour in the CNS dataset: Females display higher level of routine diversity and propensity towards exploration, in accordance with \cite{mollgaard2017correlations}. Time coverage, measuring the fraction of time an individual position is known, plays a significant role in explaining spatial entropy and activity space stability, since individuals who spend long time in the same place (or leave their phone in the same place) are more easily geo-localised. Age differences are not present within the sample of students participating in the CNS study, and they are not estimated to be relevant with respect to spatial behaviour in the MDC study. 

\textbf{We do not identify distinct classes of individuals.} A natural question is whether or not, in the samples considered, there is evidence for distinct classes of individuals based on their socio-spatial behaviour \cite{pappalardo2015returners, miritello2013limited}. We approach this problem by reducing the set of  metrics to a smaller number of uncorrelated variables by applying Principal Component Analysis \cite{wold1987principal,mika1999kernel}. The principal components represent the data through linear combinations of the original variables: In Table \ref{table_variance} we report the percentage of variance in the data explained by all components; in Table \ref{table_components} we report the coefficients $w$ describing how the original variables are linearly combined to obtain the first two principal components.  

In both datasets, we find that the first principal component (PC 0) explains $\sim 40\%$ of the differences between individuals (see Table \ref{table_variance}). For the CNS dataset, the variables contributing the most to PC 0 (e.g. such that $w^2>0.1$) are, in order, the activity space size $C$, the social circle size $k$, the number of new locations/week $n_{loc}$, the activity space entropy $H_{AS}$ and the number of new ties/week $n_{tie}$. $n_{loc}$ and $n_{tie}$ characterise the attitude towards exploration. The other metrics ($C$, $k$ and $H_{AS}$) are related to routine diversity, or the tendency to dispose of a large set of familiar locations and friends. Since the sign of $w$ is the same for all the metrics above, we can conclude that individuals with higher propensity towards exploration tend to have a more diverse social and spatial routine, and vice-versa. Similar conclusions could be drawn by looking at results obtained for the MDC dataset.

\begin{table}[h!]
\centering
\begin{tabularx}{\columnwidth}{@{}X rrrrrrrrrr @{}}
{} &  PC 0 &  PC 1 &  PC 2 &  PC 3 &  PC 4 &  PC 5 &  PC 6 &  PC 7 &  PC 8 &  PC 9 \\
\midrule
CNS &  0.39 &  0.17 &  0.12 &  0.08 &  0.07 &  0.06 &  0.04 &  0.03 &  0.03 &  0.01 \\
MDC &  0.43 &  0.14 &  0.13 &  0.08 &  0.07 &  0.06 &  0.04 &  0.03 &  0.02 &  0.01 \\
\bottomrule
\end{tabularx}
\caption{\textbf{Variance explained by principal components.} The fraction of variance explained by each principal component for the CNS and MDC dataset. }
\label{table_variance}
\end{table}

\begin{table}[h!]
\centering
\begin{tabularx}{\columnwidth}{@{} X rr rr@{}}
{} & \multicolumn{2}{ c }{CNS} & \multicolumn{2}{ c }{MDC} \\
\cmidrule(lr){2-3}
\cmidrule(lr){4-5}
{} &  $w$ (PC 0) &  $w$ (PC 1) & $w$ (PC 0) & $w$ (PC 1) \\
\midrule
Social circle size, $k$                &  0.41 &  0.16 &  0.37 & -0.15 \\
Activity space size, $C$               &  0.42 & -0.24 &  0.42 & -0.08 \\
New ties/week, $n_{tie}$          &  0.33 &  0.28 &  0.27 &  0.33 \\
New locations/week, $n_{loc}$               &  0.38 & -0.05 &  0.37 &  0.19 \\
Social circle entropy, $H_{SC}$        &  0.31 &  0.30 &  0.34 &  0.09 \\
Activity space entropy, $H_{AS}$       &  0.38 & -0.16 &  0.30 & -0.07 \\
Social circle stability, $J_{SC}$      & -0.16 & -0.46 &  0.07 & -0.72 \\
Activity space stability, $J_{AS}$     & -0.10 & -0.49 & -0.12 & -0.51 \\
Social circle rank turnover, $R_{SC}$  & -0.20 &  0.28 & -0.33 &  0.10 \\
Activity space rank turnover, $R_{AS}$ & -0.30 &  0.44 & -0.38 &  0.17 \\
\bottomrule
\end{tabularx}
\caption{\textbf{Contribution of the original variables to the first principal component.} The scalar projection $w$ of the first (PC 0) and second (PC 1) principal components, along the axis defined by each of the original variables. Each principal component has unit norm, hence the sum of $w^2$ is $1$. Results are shown for the CNS and MDC datasets. }
\label{table_components}
\end{table}

The second principal component (PC 1) accounts for $\sim 15\%$ of the total variation (see Table \ref{table_variance}). It is dominated by the social circle stability $J_{SC}$ (CNS: $w^2 =0.21$, MDC: $w^2 = 0.52$) and the activity space stability $J_{AS}$ (CNS: $w^2 = 0.24$, MDC: $w^2  =0.26$) for both datasets (see Table \ref{table_components}). The sign of the coefficients $w$ for $J_{SC}$ and $J_{AS}$ are the same, further confirming that these two metrics are correlated (see also Fig. \ref{social vs spatial_CNS}). We can conclude that the second principal component accounts for the effects of routine evolution, or the tendency to change familiar locations and friends over long time scales. We consider the first two principal components, PC 0 and PC 1, to reduce the effects of noise and we test the hypothesis that there exists different classes of individuals applying the gap statistic method \cite{tibshirani2001estimating}.
We apply it by looking at the gap between the within-cluster dispersion expected under a uniform distribution of the data and the dispersion obtained after applying K-means. For all possible choices of $K>1$, we find that the gap is not large enough to support the existence of more than one class of individuals.

\textbf{The big-five personality traits partly explain spatial and social behaviour.} We verify if the differences between individuals can be explained by the Big five personality traits model \cite{john1999big}, typically used to describe social and emotional life (see Table \ref{table_traits}). We build two multiple linear regression models that use the Big five personality traits as regressors and one of the principal components describing socio-spatial behaviour as target. Results, shown in Table \ref{table_traits_model}, show that three personality traits, neuroticism, openness and extraversion, are relevant predictors for socio-spatial behaviour. In particular, extraversion is the most important predictor of the first principal component: it positively correlates with the tendency to diversify routine and to explore opportunities. Neuroticism and openness explain instead the second principal component, since it correlates with the tendency to change routine over time (see also Fig.~\ref{personality_socio_spatial}). 

Finally, we perform all analyses considering only spatial metrics. Results are in line with those obtained considering all metrics: The first two principal components account for a large fraction of the variability in the data (see Table \ref{table_variance_2}); The first component is dominated by the activity space size $C$, the number of new locations/week $n_{loc}$ and the activity space entropy $H_{AS}$, while the second is mostly controlled by the activity space stability $J_{AS}$ (Table \ref{table_components_2}). For the CNS dataset, extraversion is the most important predictor of the first principal component, while openness, extraversion and neuroticism account for the second component (see Table \ref{table_traits_model_2} and Fig.~\ref{personality_spatial}). The result presented above hold when choosing a time-window with length $T=30$ weeks (see Supplementary Material, section 2).

\begin{table}
\centering
\begin{tabularx}{\columnwidth}{@{} X l @{}}

{Trait} & {Related Adjectives} \\
\midrule
Extraversion &  Active, Assertive, Energetic, Enthusiastic, Outgoing, Talkative\\
Agreeableness  & Appreciative, Forgiving, Generous, Kind, Sympathetic\\
Conscientiousness  & Efficient, Organised, Planful, Reliable, Responsible, Thorough\\
Neuroticism  & Anxious, Self-pitying, Tense, Touchy, Unstable, Worrying\\
Openness to Experience & Artistic, Curious, Imaginative, Insightful, Original, Wide Interests\\
\bottomrule
\end{tabularx}
\caption{The Big-Five traits and examples of adjectives describing them \cite{mccrae1992introduction}}
\label{table_traits}
\end{table}

\begin{table}[h!]
\centering
\begin{tabularx}{\columnwidth}{@{} X rrr rrr@{}}
 & \multicolumn{3}{r}{\makecell[c]{PC 0\\ $R^2=0.17$, $F=21.40$, $p_F=0.0$}} & \multicolumn{3}{r}{\makecell[c]{PC 1\\ $R^2=0.03$, $F=3.64$, $p_F=0.0$}} \\
 \cmidrule(lr){2-4}
 \cmidrule(lr){5-7}
{} &                                              coeff &        p val &   LMG &                                             coeff & p val &   LMG \\
\hline
extraversion      &                                    $0.85 \pm 0.09$ &  $<10^{-19}$ &  0.85 &                                   $0.12 \pm 0.06$ &  0.05 &  0.14 \\
openness          &                                   $-0.17 \pm 0.08$ &         0.03 &  0.02 &                                   $0.13 \pm 0.06$ &  0.02 &  0.33 \\
neuroticism       &                                    $0.25 \pm 0.09$ &        0.004 &  0.04 &                                   $0.15 \pm 0.06$ &  0.02 &   0.3 \\
agreeableness     &                                    $0.11 \pm 0.08$ &          0.2 &  0.04 &                                  $-0.07 \pm 0.06$ &   0.2 &  0.12 \\
conscientiousness &                                    $0.06 \pm 0.08$ &          0.4 &  0.04 &                                  $-0.07 \pm 0.06$ &   0.2 &  0.11 \\
\hline
\end{tabularx}
\caption{\textbf{Extraversion, openness, and neuroticism explain socio-spatial behaviour.} The result of a multiple linear regression explaining principal components of socio-spatial data (see Table \ref{table_components}). The value of each coefficient (coeff) is reported together with the probability (p val) that the coefficient is not relevant for the model. The relative importance of each coefficient (LMG) is computed using the LMG method \cite{gromping2006relative}. }
\label{table_traits_model}
\end{table}

\begin{figure}
\includegraphics[width=.9\textwidth]{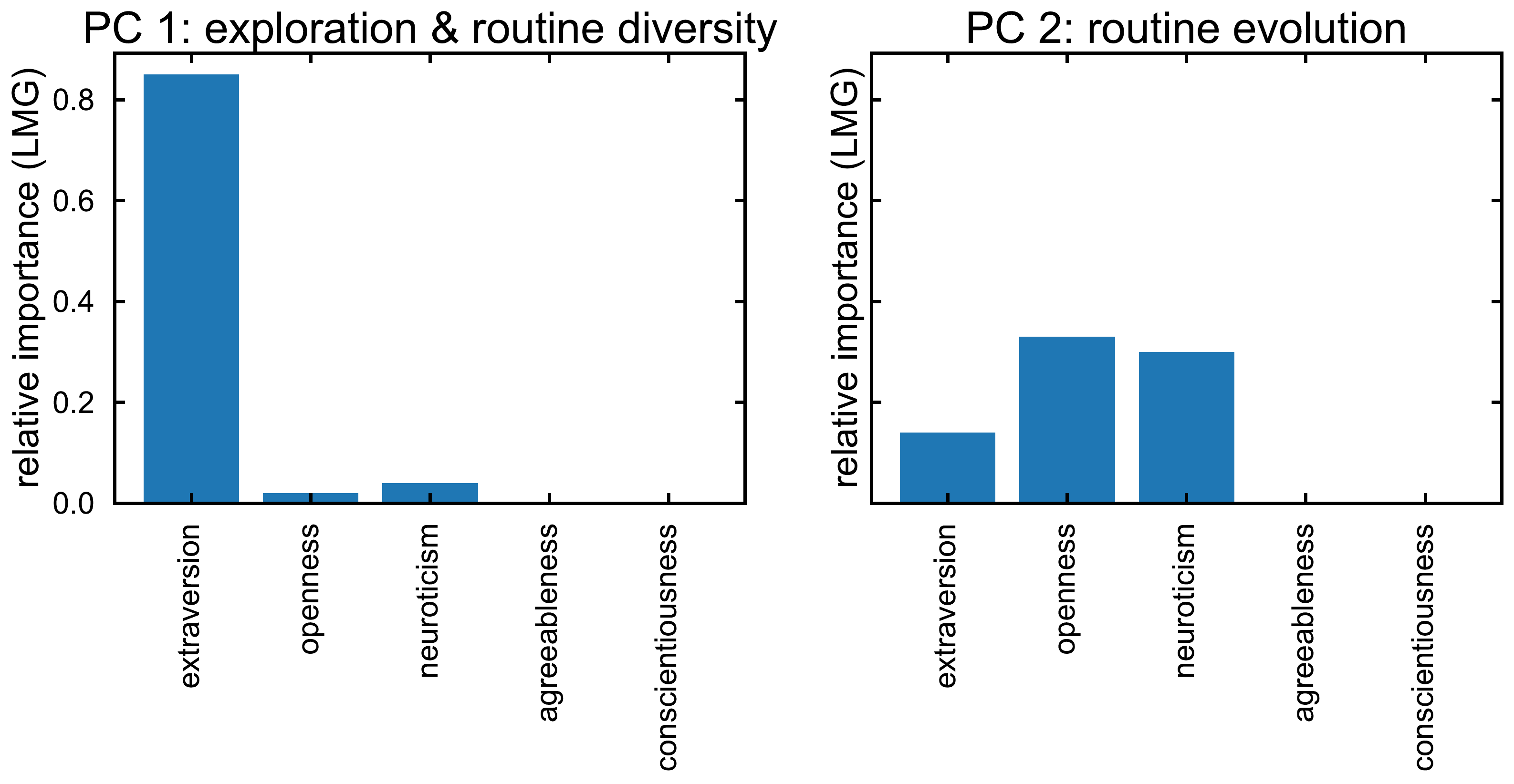}
\caption{\textbf{Relative importance of personality traits for socio-spatial behaviour} LMG of each regressor computed using the  Lindeman, Merenda and Gold method \cite{gromping2006relative} for the multiple regression model of the principal components (see also Table \ref{table_traits_model}).}
\label{personality_socio_spatial}
\end{figure}

\begin{figure}
\includegraphics[width=.9\textwidth]{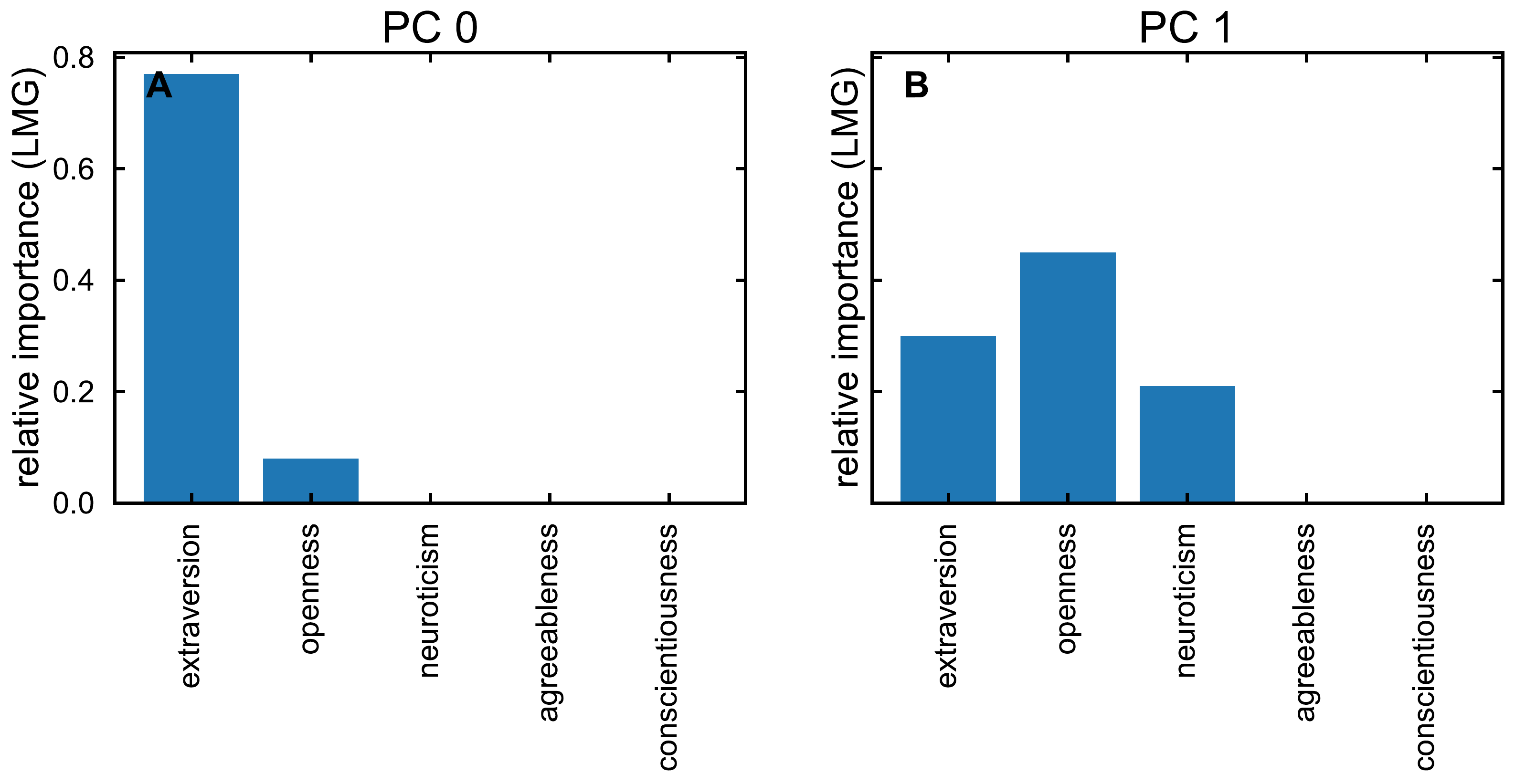}
\caption{\textbf{Relative importance of personality traits for spatial behaviour} LMG of each regressor computed using the  Lindeman, Merenda and Gold method \cite{gromping2006relative} for the multiple regression model of the principal components (see also Table \ref{table_traits_model_2}).}
\label{personality_spatial}
\end{figure}

\begin{table}[h!]
\centering
\begin{tabularx}{\columnwidth}{@{}X rrrrr@{}}
{} &  PC 0 &  PC 1 &  PC 2 &  PC 3 &  PC 4 \\
\midrule
CNS &  0.53 &  0.21 &  0.13 &  0.10 &  0.04 \\
MDC &  0.56 &  0.19 &  0.13 &  0.07 &  0.04 \\
\bottomrule
\end{tabularx}
\caption{\textbf{Variance explained by principal components (only spatial data).} The fraction of variance explained by each principal component for the CNS and MDC dataset. }
\label{table_variance_2}
\end{table}

\begin{table}[h!]
\centering
\begin{tabularx}{\columnwidth}{@{} X rr rr @{}}
{} & \multicolumn{2}{c}{CNS} & \multicolumn{2}{c}{MDC} \\
\cmidrule(lr){2-3}
\cmidrule(lr){4-5}
{} &  PC 0 &  PC 1 &  PC 0 &  PC 1 \\
\midrule
Activity space size, $C$               & -0.58 &  0.02 &  0.55 &  0.12 \\
New locations/week, $n_{loc}$               & -0.48 & -0.19 &  0.51 & -0.09 \\
Activity space entropy, $H_{AS}$       & -0.50 & -0.08 &  0.43 &  0.20 \\
Activity space stability, $J_{AS}$     & -0.02 &  0.94 & -0.19 &  0.95 \\
Activity space rank turnover, $R_{AS}$ &  0.43 & -0.25 & -0.47 & -0.16 \\
\bottomrule
\end{tabularx}\caption{\textbf{Principal Components (only spatial data).} The weight of each metric in the first two principal components, for both datasets. }
\label{table_components_2}
\end{table}

\begin{table}[h!]
\centering
\begin{tabularx}{\columnwidth}{@{}X  lll lll @{}}
 & \multicolumn{3}{c}{\makecell[c]{PC 0\\ $R^2=0.10$, $F=12.83$, $p_F=0.0$}} & \multicolumn{3}{ c}{\makecell[c]{PC 1\\ $R^2=0.03$, $F=3.50$, $p_F=0.0$}} \\
\cmidrule(lr){2-4}
\cmidrule(lr){5-7}
{} &                                              coeff &        p val &   LMG &                                             coeff &  p val &   LMG \\
\hline
extraversion      &                                   $-0.50 \pm 0.07$ &  $<10^{-10}$ &  0.77 &                                  $-0.11 \pm 0.05$ &   0.02 &   0.3 \\
openness          &                                    $0.19 \pm 0.07$ &        0.004 &  0.08 &                                  $-0.11 \pm 0.04$ &  0.009 &  0.45 \\
neuroticism       &                                   $-0.07 \pm 0.07$ &          0.4 &  0.03 &                                  $-0.10 \pm 0.05$ &   0.03 &  0.21 \\
agreeableness     &                                   $-0.10 \pm 0.07$ &          0.2 &  0.07 &                                   $0.01 \pm 0.05$ &    0.8 &  0.01 \\
conscientiousness &                                   $-0.05 \pm 0.07$ &          0.5 &  0.05 &                                   $0.03 \pm 0.04$ &    0.4 &  0.03 \\
\hline
\end{tabularx}
\caption{\textbf{Extraversion, openness, and neuroticism explain spatial behaviour.} The result of a multiple linear regression explaining principal components of spatial data (see Table \ref{table_components}). The value of each coefficient (coeff) is reported together with the probability (p val) that the coefficient is not relevant for the model. The relative importance of each coefficient (LMG) is computed using the LMG method \cite{gromping2006relative}. }
\label{table_traits_model_2}
\end{table}

\section*{Discussion}

Using high resolution data from two large scale studies, we have investigated the connection between social and spatial behaviour for the first time. We have shown that, in both domains, individuals balance the trade-off between exploring new opportunities and exploiting known options in a distinctive and persistent manner. We have found that, to a significant extent, individuals adopt a similar strategy in the social and spatial sphere. These strategies are heterogeneous across the two samples considered, and there is no evidence suggesting that there exist distinct classes of individuals. Finally, we have shown that the big five personality traits explain related aspects of both social and spatial behaviour. In particular, we have found that extraverted individuals are more explorative and have diverse routines in both the social and the spatial sphere while neuroticism and openness associate with high level of routine instability in the social and spatial domain.

Our findings confirm the usefulness of mobile phone data to study the connections between behaviour and personality \cite{lambiotte2014tracking, bogomolov2014daily, staiano2012friends,centellegher2017personality,de2013predicting,chittaranjan2013mining}.
The results are in line with previous findings on the relation between personality and social behaviour: extraversion correlates with ego-network size \cite{pollet2011extraverts, asendorpf2003personality, casciaro1998seeing} and diverse composition \cite{friggeri2012psychological}, openness to experience to social network turnover \cite{centellegher2017personality} and neuroticism does not correlate with social network size \cite{roberts2008individual}. Furthermore, our findings establish a relation between personality and spatial behaviour, validating the theories suggesting that spatial choices are partially dictated by personality dispositions \cite{aitken1991person} and that a single set of personality traits underlies many aspects of a person's behaviour \cite{mischel1973toward,allport1937personality}. 

Our findings on the connection between spatial behaviour and personality are consistent with the existing literature on personality. The correlation between exploration and extraversion could be explained by the fact that extraverted individuals are more likely to be risk-takers in various domains of life \cite{nicholson2005personality}. Extraverted individuals are also generally more likely to engage in social activities \cite{lucas2000cross}, which could partially explain why they allocate time among a larger set of locations. Furthermore, the key finding that individuals who score high in neuroticism and openness display a tendency to change familiar locations, and friends, over time fits well within the existing picture. In the case of neuroticism, it is well known that this trait is closely related with `stability' \cite{robinson2005neuroticism}, such that the trait of neuroticism is sometimes referred to as (low) `emotional stability' \cite{judge2001relationship}.
Also, at the core of neuroticism is the tendency to experience negative emotions \cite{costa1992revised} including dissatisfaction \cite{costa1980influence}, which in turn can lead into desire for change \cite{zhou2001job,rusbult1983responses}.
Finally, it is known that people scoring high in neuroticism have a larger number of weak ties \cite{kalish2006psychological} and perceive that they tend to have less
social support \cite{russell1997personality,stokes1985relation}, in line with our observation that they dispose of an unstable ego-network. Openness to experience has been shown to correlate with `disloyal' behaviour also in other contexts such as politics \cite{bakker2016stay} and shopping \cite{matzler2006individual}.  
Our results, in agreement with previous studies on social \cite{lambiotte2014tracking, bogomolov2014daily, staiano2012friends,centellegher2017personality,de2013predicting,chittaranjan2013mining} and online \cite{bachrach2012personality,markovikj2013mining,quercia2011our,kosinski2013private} 
behaviour, show that personality traits explain only partially how individuals behave in specific situations \cite{fleeson2008end}.

As a final point, we emphasize that the individual characterisation of spatial behaviour and connections with personality are fundamental to develop conceptual \cite{van2010transport} and predictive \cite{jiang2016timegeo} models of travel behaviour accounting for individual-level differences. 

\clearpage


\section*{Competing interests}
  The authors declare that they have no competing interests.

\section*{Author's contributions}
    LA, SL and AB designed the study; LA performed the analysis; LA, SL and AB analysed the results; LA, SL and AB wrote the paper. 

\section*{Acknowledgements}
This work was supported by the Danish Council for Independent Research (``Microdynamics of influence in social systems", grant id. 4184-00556, SL is PI). Portions of the research in this paper used the MDC Database made available by Idiap Research Institute, Switzerland and owned by Nokia. 

\section*{List of Abbreviations}
CNS: Copenhagen Networks Study \\
MDC: (Lausanne) Mobile Data Challenge \\ 
GSM: Global System for Mobile Communications \\
JSD: Jensen-Shannon divergence \\ 
LMG: Lindeman, Merenda and Gold \\

\clearpage
\pagebreak

\bibliographystyle{unsrt}
\bibliography{personality_biblio}

\clearpage

\begin{center}
{\LARGE Supplementary Material for \\ { Understanding the interplay between social and spatial behaviour}}\\[0.7cm]
\end{center}

\beginsupplement
\section{Results obtained with the MDC dataset}
\begin{figure*}[h!]
\centering
\includegraphics[width=.9\textwidth]{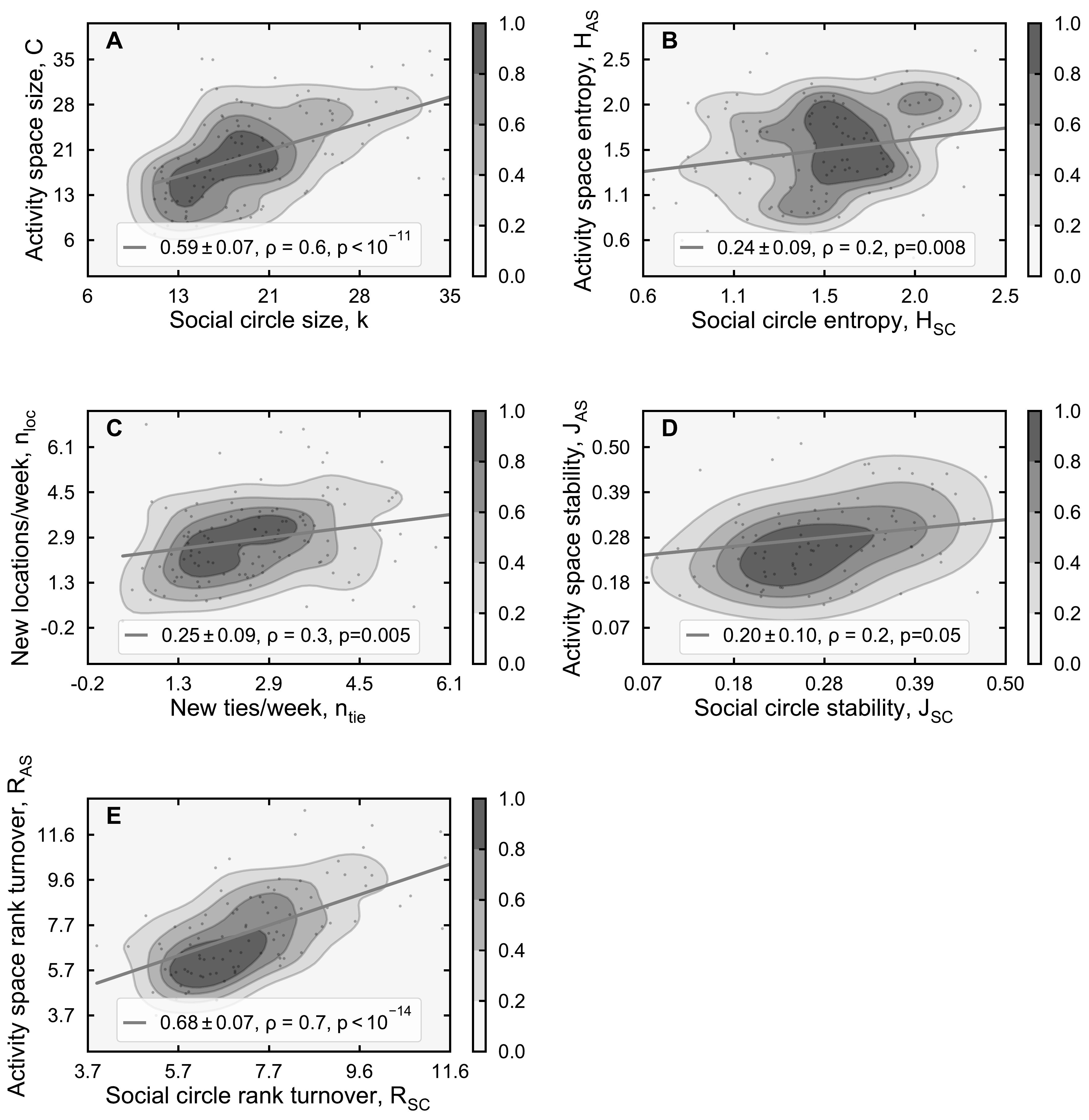}
\caption{\textbf{MDC dataset: correlation between the four dimensions of social and spatial behaviour.} ({A}) Activity space vs social circle size.  ({B}) Activity space vs social circle composition measured as their entropy. ({C}) Average number of new locations vs new ties per week. ({D}) Stability of the activity space vs the stability of the social circle measured as the Jaccard similarity between their composition in consecutive time-windows. ({E}) Rank turnover of the activity space vs the rank turnover of the social circle. Coloured filled areas correspond to cumulative probabilities estimated via Gaussian Kernel Density estimations. Grey lines correspond to linear fit with angular coefficient $b$ reported in the legend. The Pearson correlation coefficient, with corresponding p-value, is reported in the legend.}
\label{social vs spatial_MDC}
\end{figure*}

Tables \ref{table1_MDC}, \ref{multiple regression MDC} and Fig.~\ref{social vs spatial_MDC} report the results of the persistence analysis, the multiple regression analysis, and the correlation analysis for the MDC dataset.

\renewcommand*{\arraystretch}{1.5}

\begin{table}[h!]
\centering
\begin{tabularx}{\columnwidth}{@{}X rrr@{}}
{} & $\overline{d_{self}}$ & $\overline{d_{ref}}$ & $\overline{d_{self}(i)<d_{ref}(i,j)}$ \\
\midrule
Social circle size, $k$                &       $0.05 \pm 0.13$ &           $10 \pm 5$ &                               $97 $\% \\
Activity space size, $C$               &       $0.07 \pm 0.12$ &            $8 \pm 3$ &                               $97 $\% \\
New ties/week, $n_{tie}$          &         $0.2 \pm 0.3$ &            $2 \pm 1$ &                               $91 $\% \\
New locations/week, $n_{loc}$               &         $0.2 \pm 0.6$ &            $2 \pm 1$ &                               $90 $\% \\
Social circle entropy, $H_{SC}$        &     $0.006 \pm 0.014$ &        $0.7 \pm 0.3$ &                               $97 $\% \\
Activity space entropy, $H_{AS}$       &     $0.004 \pm 0.008$ &        $0.5 \pm 0.2$ &                               $97 $\% \\
Social circle stability, $J_{SC}$      &     $0.002 \pm 0.005$ &      $0.15 \pm 0.05$ &                               $99 $\% \\
Activity space stability, $J_{AS}$     &     $0.002 \pm 0.004$ &      $0.12 \pm 0.05$ &                               $99 $\% \\
Social circle rank turnover, $R_{SC}$  &       $0.07 \pm 0.15$ &            $2 \pm 1$ &                               $98 $\% \\
Activity space rank turnover, $R_{AS}$ &         $0.2 \pm 0.6$ &            $2 \pm 1$ &                               $97 $\% \\
\bottomrule
\end{tabularx}
\caption{\textbf{MDC dataset: Persistence of social and spatial behaviour. } For each of the social and spatial metrics, $\overline{d_{self}}$ is the average self-distance and $\overline{d_{ref}}$ is the reference distance between an individual and all others, averaged across individuals. The third column reports the fraction of cases where $\overline{d_{self}(i)<d_{ref}(i,j)}$, averaged across the population. }
\label{table1_MDC}
\end{table}

\def\arraystretch{1.5}
\begin{table}
\begin{tabularx}{\columnwidth}{@{} X rrr rrr @{}}
\hline
 \textbf{Model M1:  Activity space size, $C$} &                                             coeff &        p val &   LMG \\
\hline
Social circle size, $k$ &                                         $5 \pm 1$ &  $<10^{-11}$ &  0.98 \\
gender                  &                                     $0.1 \pm 0.6$ &          0.8 &  0.01 \\
age group               &                                     $0.6 \pm 0.6$ &          0.3 &  0.01 \\
time coverage           &                                    $-0.4 \pm 0.6$ &          0.4 &   0.0 \\
{[$R^2=0.40$, $F=16.80$, $p_F=0.0$]}
\\
\\
 \textbf{Model M2:  Activity space entropy, $H_{AS}$}  &                                             &  &   \\
\hline
Social circle entropy, $H_{SC}$ &                                  $0.11 \pm 0.04$ &  0.009 &  0.28 \\
gender                          &                                  $0.04 \pm 0.04$ &    0.3 &  0.03 \\
age group                       &                                 $-0.08 \pm 0.04$ &   0.06 &  0.21 \\
time coverage                   &                                 $-0.14 \pm 0.04$ &  0.002 &  0.48 \\
{[$R^2=0.20$, $F=6.50$, $p_F=0.0$]}
\\
\\
 \textbf{Model M3:  New locations/week, $n_{loc}$}  &                                             &  &    \\
\hline
New ties/week, $n_{tie}$ &                                    $0.5 \pm 0.1$ &  0.002 &  0.69 \\
gender                        &                                  $0.01 \pm 0.15$ &    0.9 &  0.04 \\
age group                     &                                    $0.2 \pm 0.1$ &    0.2 &   0.1 \\
time coverage                 &                                   $-0.3 \pm 0.1$ &   0.06 &  0.17 \\
{[$R^2=0.13$, $F=3.78$, $p_F=0.0$]}
\\
\\
 \textbf{Model M4: Activity space stability, $J_{AS}$} &                                             & &    \\
\hline
Social circle stability, $J_{SC}$ &                                  $0.02 \pm 0.01$ &   0.1 &  0.82 \\
gender                            &                               $-0.006 \pm 0.012$ &   0.6 &  0.15 \\
age group                         &                               $-0.003 \pm 0.012$ &   0.8 &  0.03 \\
time coverage                     &                $({-10} \pm {1213}) \cdot10^{-5}$ &   1.0 &   0.0 \\
{[$R^2=0.04$, $F=0.80$, $p_F=0.5$]}
\\
\\
 \textbf{Model M5:  Activity space rank turnover, $R_{AS}$} &                                              &      &   \\
\hline
Social circle rank turnover, $R_{SC}$ &                                         $1 \pm 0$ &  $<10^{-15}$ &  0.97 \\
gender                                &                                   $0.04 \pm 0.15$ &          0.8 &  0.02 \\
age group                             &                                    $-0.2 \pm 0.1$ &          0.1 &  0.01 \\
time coverage                         &                                  $-0.06 \pm 0.15$ &          0.7 &   0.0 \\
{[$R^2=0.55$, $F=27.24$, $p_F=0.0$]}
\\
\\
\hline
\end{tabularx}

\caption{\textbf{Linear regression models for the MDC dataset.} For each model, we report the $R^2$ goodness of fit, the $F-test$ statistics with the corresponfing p-value $p_F$. We show the coefficients (coeff) calculated by the regression model, the probability (p val) that the variable is not relevant, and the relative importance (LMG) of each regressor computed using the  Lindeman, Merenda and Gold method. Gender is a binary variable taking value 1 for females and 2 for males.}
\label{multiple regression MDC}
\end{table}

\clearpage
\section{Results obtained with other windows}
Figs.~\ref{social vs spatial_CNS}, \ref{personality_socio_spatial}, \ref{personality_socio_spatial}, \ref{personality_spatial} and Tables \ref{table1_CNS}, \ref{multiple regression CNS}, \ref{table4}, \ref{table5}, \ref{table7}, \ref{table9}, \ref{table10}, \ref{table11} report the results obtained choosing a time-window with length $T=30$ weeks (see main manuscript, section `Methods').

\begin{figure*}[h!]
\centering
\includegraphics[width=.9\textwidth]{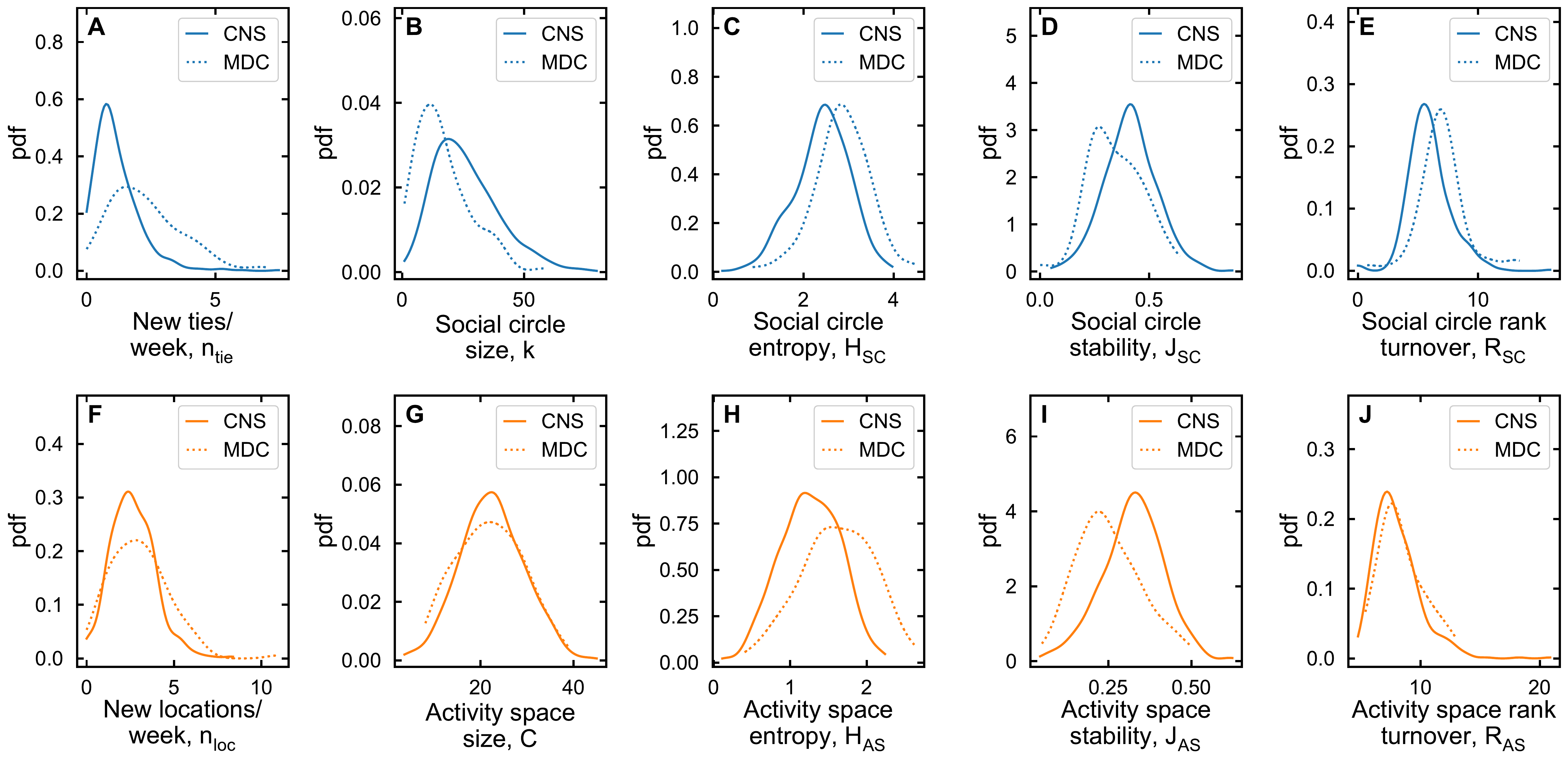}
\caption{\textbf{T=30, Distribution of social (above line) and spatial (bottom line) metrics for the CNS and MDC datasets.}}
\label{distributions}
\end{figure*}

\begin{figure*}[h!]
\centering
\includegraphics[width=.9\textwidth]{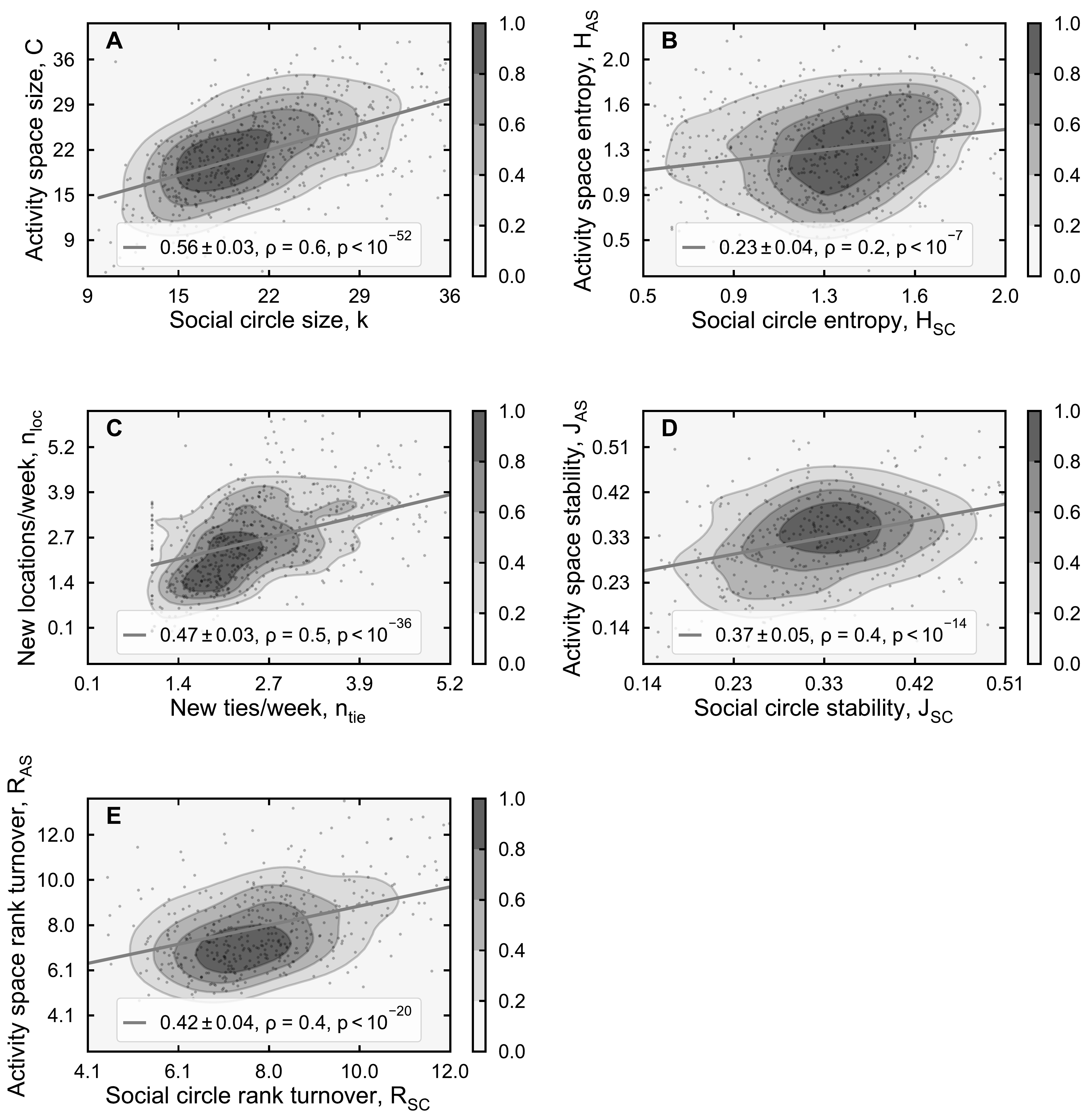}
\caption{\textbf{T=30, CNS dataset: correlation between the four dimensions of social and spatial behaviour.} ({A}) Activity space vs social circle size.  ({B}) Activity space vs social circle composition measured as their entropy. ({C}) Average number of new locations vs new ties per week. ({D}) Stability of the activity space vs the stability of the social circle measured as the Jaccard similarity between their composition in consecutive time-windows. ({E}) Rank turnover of the activity space vs the rank turnover of the social circle. Coloured filled areas correspond to cumulative probabilities estimated via Gaussian Kernel Density estimations. Grey lines correspond to linear fit with angular coefficient $b$ reported in the legend. The Pearson correlation coefficient, with corresponding p-value, is reported in the legend. }
\label{social vs spatial_CNS}
\end{figure*}

\begin{figure}
\includegraphics[width=\textwidth]{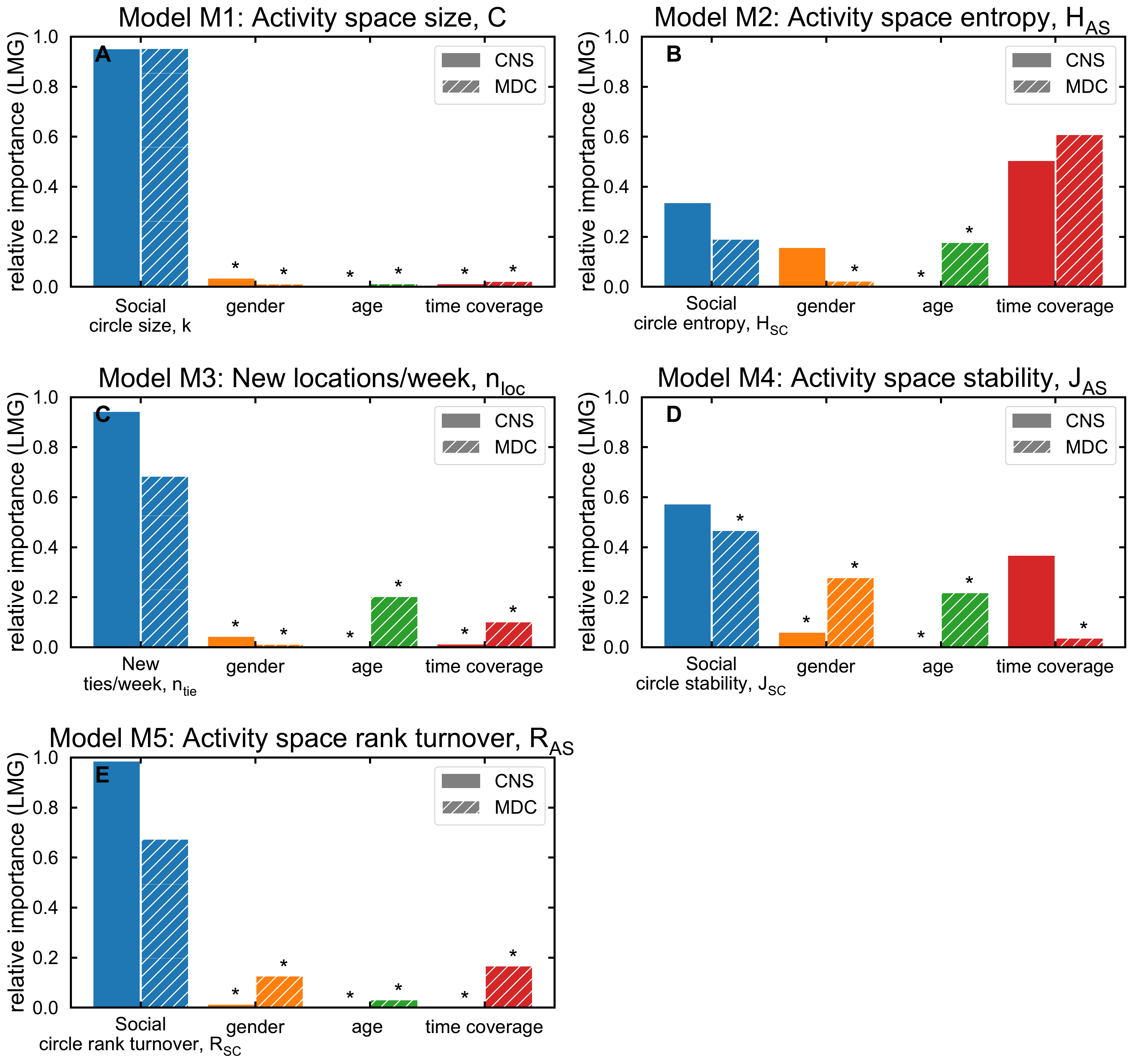}
\caption{\textbf{T=30, Relative importance of regressors} LMG of each regressor computed using the  Lindeman, Merenda and Gold method for models M1 (A), M2 (B), M3 (C), M4 (D) and M5 (E). Plain bars show results for the CNS dataset, dashed bars for the MDC dataset. Variables that are not significant in the regression model are marked with *.  }
\label{multiple regression_figure}
\end{figure}

\begin{figure}
\includegraphics[width=.9\textwidth]{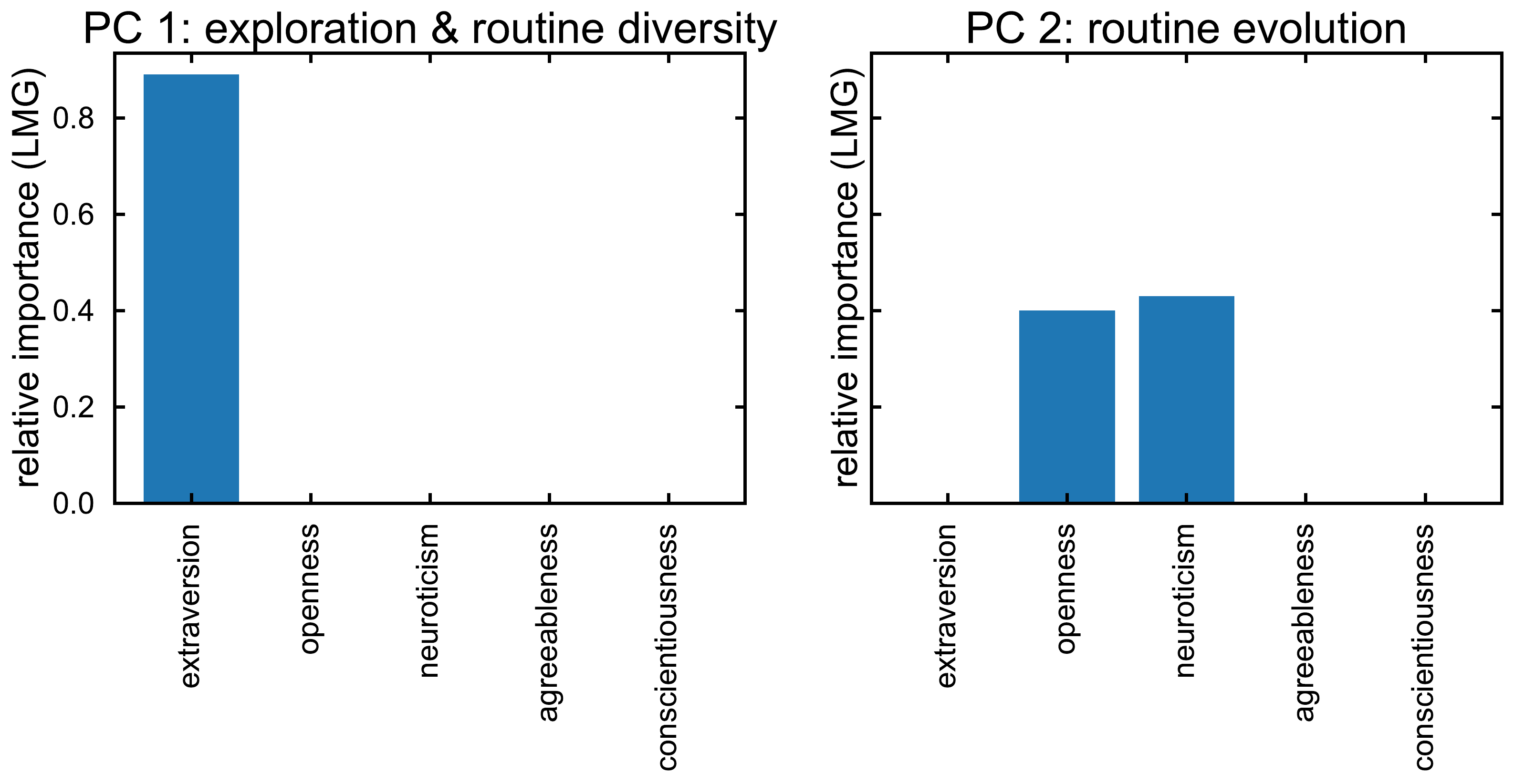}
\caption{\textbf{T=30, Relative importance of personality traits for socio-spatial behaviour} LMG of each regressor computed using the  Lindeman, Merenda and Gold method for the multiple regression model of the principal components (table \ref{table7}).}
\label{personality_socio_spatial}
\end{figure}

\begin{figure}
\includegraphics[width=.9\textwidth]{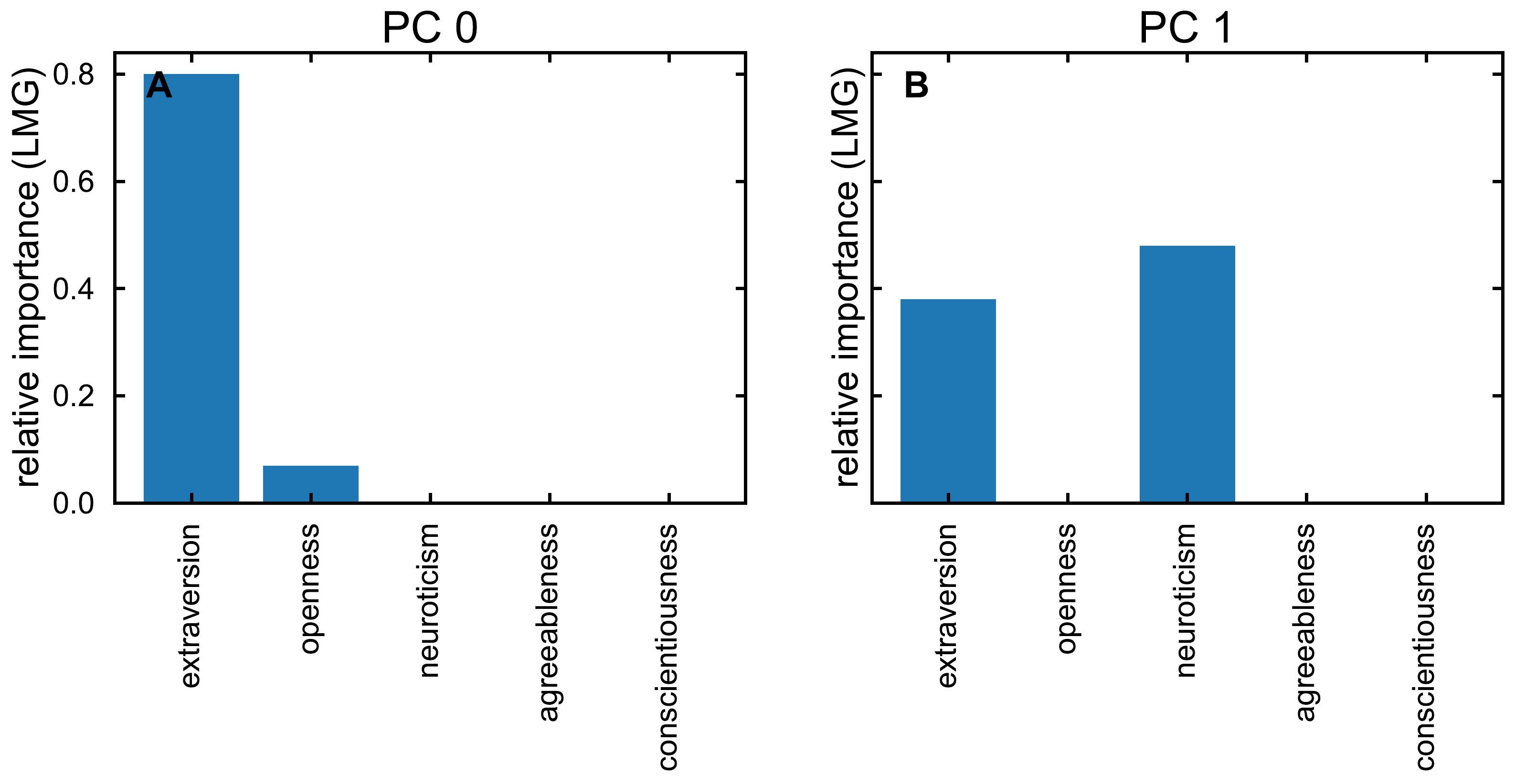}
\caption{\textbf{T=30, Relative importance of personalitry traits for spatial behaviour} LMG of each regressor computed using the  Lindeman, Merenda and Gold method for the multiple regression model of the principal components (table \ref{table11}).}
\label{personality_spatial}
\end{figure}

\clearpage

\renewcommand*{\arraystretch}{1.5}

\begin{table}[h!]
\centering
\begin{tabularx}{\columnwidth}{@{}X rrr}
{} &          $\overline{d_{self}}$ & $\overline{d_{ref}}$ & $\overline{d_{self}(i)<d_{ref}(i,j)}$ \\
\midrule

Social circle size, $k$                &                $0.04 \pm 0.13$ &           $15 \pm 6$ &                              $100 $\% \\
Activity space size, $C$               &                $0.04 \pm 0.07$ &            $8 \pm 3$ &                               $99 $\% \\
New ties/week, $n_{tie}$          &                $0.06 \pm 0.12$ &        $0.9 \pm 0.5$ &                               $96 $\% \\
New locations/week, $n_{loc}$               &                  $0.1 \pm 0.2$ &            $1 \pm 1$ &                               $95 $\% \\
Social circle entropy, $H_{SC}$        &              $0.002 \pm 0.005$ &        $0.7 \pm 0.3$ &                               $99 $\% \\
Activity space entropy, $H_{AS}$       &              $0.002 \pm 0.006$ &        $0.4 \pm 0.1$ &                               $99 $\% \\
Social circle stability, $J_{SC}$      &  $({6} \pm {15}) \cdot10^{-4}$ &      $0.14 \pm 0.05$ &                              $100 $\% \\
Activity space stability, $J_{AS}$     &  $({6} \pm {11}) \cdot10^{-4}$ &      $0.10 \pm 0.04$ &                              $100 $\% \\
Social circle rank turnover, $R_{SC}$  &                $0.04 \pm 0.11$ &            $2 \pm 1$ &                               $99 $\% \\
Activity space rank turnover, $R_{AS}$ &                $0.04 \pm 0.20$ &            $2 \pm 1$ &                               $99 $\% \\
\bottomrule
\end{tabularx}\caption{\textbf{T=30, CNS dataset: Persistence of social and spatial behaviour.}For each of the social and spatial metrics, $\overline{d_{self}}$ is the average self-distance and $\overline{d_{ref}}$ is the reference distance between an individual and all others, averaged across individuals. The third column reports the fraction of cases where $\overline{d_{self}(i)<d_{ref}(i,j)}$, averaged across the population.}
\label{table1_CNS}

\end{table}

\begin{table}
\begin{tabularx}{\columnwidth}{@{}X rrr rrr @{}}
 \textbf{Model M1:  Activity space size, $C$}  &                                             coeff &        p val &   LMG \\
\hline
Social circle size, $k$ &                                         $4 \pm 0$ &  $<10^{-46}$ &  0.95 \\
gender                  &                                    $-0.3 \pm 0.2$ &          0.2 &  0.04 \\
time coverage           &                                     $0.5 \pm 0.2$ &         0.05 &  0.01 \\
{[$R^2=0.32$, $F=91.23$, $p_F=0.0$]}
\\
\\
\textbf{Model M2: Activity space entropy, $H_{AS}$}   &                                              &   &    \\
\hline
Social circle entropy, $H_{SC}$ &                                   $0.07 \pm 0.02$ &  $<10^{-4}$ &  0.34 \\
gender                          &                                  $-0.05 \pm 0.02$ &  $<10^{-3}$ &  0.16 \\
time coverage                   &                                  $-0.09 \pm 0.02$ &  $<10^{-8}$ &  0.51 \\
{[$R^2=0.12$, $F=26.82$, $p_F=0.0$]}
\\
\\
 \textbf{Model M3:  New locations/week, $n_{loc}$} &                                              &    &   \\
\hline
New ties/week, $n_{tie}$ &                                   $0.58 \pm 0.05$ &  $<10^{-30}$ &  0.94 \\
gender                        &                                  $-0.09 \pm 0.05$ &         0.04 &  0.04 \\
time coverage                 &                                   $0.03 \pm 0.05$ &          0.5 &  0.01 \\
{[$R^2=0.22$, $F=55.56$, $p_F=0.0$]}
\\
\\
\textbf{Model M4:  Activity space stability, $J_{AS}$} &                                               &       &  \\
\hline
Social circle stability, $J_{SC}$ &                                 $0.027 \pm 0.004$ &  $<10^{-9}$ &  0.57 \\
gender                            &                                 $0.009 \pm 0.004$ &        0.02 &  0.06 \\
time coverage                     &                                 $0.020 \pm 0.004$ &  $<10^{-5}$ &  0.37 \\
{[$R^2=0.18$, $F=30.32$, $p_F=0.0$]} 
\\
\\
\textbf{Model M5: Activity space rank turnover, $R_{AS}$}  &                                              &         &    \\
\hline
Social circle rank turnover, $R_{SC}$ &                                   $0.81 \pm 0.08$ &  $<10^{-19}$ &  0.99 \\
gender                                &                                   $0.09 \pm 0.08$ &          0.3 &  0.01 \\
time coverage                         &                                $-0.001 \pm 0.084$ &          1.0 &   0.0 \\
{[$R^2=0.18$, $F=31.70$, $p_F=0.0$]}
\\
\\
\hline
\end{tabularx}
\caption{T=30, Linear regression models for the CNS dataset. For each model, we report the $R^2$ goodness of fit, the $F-test$ statistics with the corresponfing p-value $p_F$. We show the coefficients (coeff) calculated by the regression model, the probability (p val) that the variable is not relevant, and the relative importance (LMG) of each regressor computed using the  Lindeman, Merenda and Gold method. Gender is a binary variable taking value 1 for females and 2 for males. For this dataset, age is not relevant as all participants have similar age. }
\label{multiple regression CNS}
\end{table}

\begin{table}[h!]
\centering
\begin{tabularx}{\columnwidth}{@{}X rrrrrrrrrr @{}}

{} &  PC 0 &  PC 1 &  PC 2 &  PC 3 &  PC 4 &  PC 5 &  PC 6 &  PC 7 &  PC 8 &  PC 9 \\
\midrule
CNS &  0.40 &  0.18 &  0.10 &  0.07 &  0.07 &  0.06 &  0.04 &  0.03 &  0.03 &  0.02 \\
MDC &  0.39 &  0.19 &  0.12 &  0.10 &  0.06 &  0.05 &  0.05 &  0.03 &  0.02 &  0.01 \\
\bottomrule
\end{tabularx}\caption{\textbf{T=30, Variance explained by principal components.} The fraction of variance explained by each principal component for the CNS and MDC dataset. }
\label{table4}
\end{table}
\begin{table}[h!]
\centering
\begin{tabularx}{\columnwidth}{@{}X rr rr @{}}
{} & \multicolumn{2}{ c}{CNS} & \multicolumn{2}{ c}{MDC} \\
\cmidrule(lr){2-3}
\cmidrule(lr){4-5}
{} &  PC 0 &  PC 1 &  PC 0 &  PC 1 \\
Social circle size, $k$                &  0.41 &  0.18 & -0.36 &  0.04 \\
Activity space size, $C$               &  0.42 & -0.23 & -0.40 & -0.05 \\
New ties/week, $n_{tie}$          &  0.33 &  0.27 & -0.24 & -0.35 \\
New locations/week, $n_{loc}$               &  0.39 & -0.11 & -0.37 & -0.22 \\
Social circle entropy, $H_{SC}$        &  0.29 &  0.33 & -0.36 & -0.23 \\
Activity space entropy, $H_{AS}$       &  0.38 & -0.10 & -0.35 &  0.13 \\
Social circle stability, $J_{SC}$      & -0.12 & -0.50 & -0.11 &  0.56 \\
Activity space stability, $J_{AS}$     & -0.06 & -0.50 & -0.03 &  0.62 \\
Social circle rank turnover, $R_{SC}$  & -0.17 &  0.26 &  0.28 & -0.19 \\
Activity space rank turnover, $R_{AS}$ & -0.35 &  0.37 &  0.42 & -0.18 \\
\bottomrule
\end{tabularx}\caption{\textbf{T=30, Principal Components.} The weight of each metric in the first two principal components, for both datasets. }
\label{table5}
\end{table}

\begin{table}[h!]
\centering
\begin{tabularx}{\columnwidth}{@{}X rrr rrr @{}}
 & \multicolumn{3}{c}{\makecell[c]{PC 0 \\ $R^2=0.17$, $F=17.08$, $p_F=0.0$}} & \multicolumn{3}{c}{\makecell[c]{PC 1\\ $R^2=0.02$, $F=2.05$, $p_F=0.1$}} \\
\cmidrule(lr){2-4}
\cmidrule(lr){5-7}
{} &                                              coeff &        p val &   LMG &                                             coeff & p val &   LMG \\
\hline
extraversion      &                                      $0.9 \pm 0.1$ &  $<10^{-15}$ &  0.89 &                                   $0.06 \pm 0.07$ &   0.4 &  0.04 \\
openness          &                                   $-0.18 \pm 0.09$ &         0.06 &  0.02 &                                   $0.13 \pm 0.07$ &  0.05 &   0.4 \\
neuroticism       &                                      $0.1 \pm 0.1$ &          0.1 &  0.03 &                                   $0.15 \pm 0.07$ &  0.04 &  0.43 \\
agreeableness     &                                    $0.05 \pm 0.10$ &          0.6 &  0.02 &                                  $-0.04 \pm 0.07$ &   0.5 &  0.09 \\
conscientiousness &                                    $0.04 \pm 0.10$ &          0.7 &  0.04 &                                  $-0.03 \pm 0.07$ &   0.7 &  0.04 \\
\hline
\end{tabularx}
\caption{\textbf{T=30, Extraversion, openness, and neuroticism explain socio-spatial behaviour.} The result of a multiple linear regression explaining principal components of socio-spatial data (Table \ref{table5}). The value of each coefficient (coeff) is reported together with the probability (p val) that the coefficient is not relevant for the model. The relative importance of each coefficient (LMG) is computed using the LMG method. }
\label{table7}
\end{table}

\begin{table}[h!]
\centering
\begin{tabularx}{\columnwidth}{@{}X  rrrrr @{}}
{} &  PC 0 &  PC 1 &  PC 2 &  PC 3 &  PC 4 \\
\midrule
CNS &  0.57 &  0.21 &  0.10 &  0.08 &  0.04 \\
MDC &  0.55 &  0.24 &  0.12 &  0.06 &  0.03 \\
\bottomrule
\end{tabularx}\caption{\textbf{T=30, Variance explained by principal components (only spatial data).} The fraction of variance explained by each principal component for the CNS and MDC dataset. }
\label{table9}
\end{table}
\begin{table}[h!]
\centering
\begin{tabularx}{\columnwidth}{@{}X rr rr @{}}
{} & \multicolumn{2}{ c}{CNS} & \multicolumn{2}{ c}{MDC} \\
\cmidrule(lr){2-3}
\cmidrule(lr){4-5}
{} &  PC 0 &  PC 1 &  PC 0 &  PC 1 \\
\hline
Activity space size, $C$               & -0.55 &  0.01 & -0.55 & -0.10 \\
New locations/week, $n_{loc}$               & -0.48 & -0.16 & -0.48 & -0.35 \\
Activity space entropy, $H_{AS}$       & -0.48 & -0.14 & -0.44 &  0.20 \\
Activity space stability, $J_{AS}$     & -0.06 &  0.96 & -0.02 &  0.88 \\
Activity space rank turnover, $R_{AS}$ &  0.48 & -0.16 &  0.51 & -0.22 \\
\hline
\end{tabularx}\caption{\textbf{T=30, Principal Components (only spatial data).} The weight of each metric in the first two principal components, for both datasets. }
\label{table10}
\end{table}

\begin{table}[h!]
\centering
\begin{tabularx}{\columnwidth}{@{}X rrr rrr @{}}
 & \multicolumn{3}{ c}{\makecell[c] {PC 0\\ $R^2=0.11$, $F=11.55$, $p_F=0.0$}} & \multicolumn{3}{c}{\makecell[c]{PC 1\\ $R^2=0.02$, $F=2.02$, $p_F=0.1$}} \\
\cmidrule(lr){2-4}
\cmidrule(lr){5-7}
{} &                                              coeff &       p val &   LMG &                                             coeff & p val &   LMG \\
\hline
extraversion      &                                   $-0.56 \pm 0.09$ &  $<10^{-9}$ &   0.8 &                                  $-0.12 \pm 0.05$ &  0.03 &  0.38 \\
openness          &                                    $0.20 \pm 0.08$ &        0.01 &  0.07 &                                  $-0.04 \pm 0.05$ &   0.5 &  0.08 \\
neuroticism       &                                    $0.03 \pm 0.08$ &         0.7 &  0.07 &                                  $-0.14 \pm 0.05$ &  0.01 &  0.48 \\
agreeableness     &                                   $-0.05 \pm 0.08$ &         0.5 &  0.03 &                                $-0.002 \pm 0.052$ &   1.0 &  0.01 \\
conscientiousness &                                 $-0.005 \pm 0.081$ &         1.0 &  0.03 &                                  $-0.03 \pm 0.05$ &   0.6 &  0.04 \\
\hline
\end{tabularx}\caption{\textbf{T=30, Extraversion, openness, and neuroticism explain spatial behaviour.} The result of a multiple linear regression explaining principal components of spatial data (Table \ref{table5}). The value of each coefficient (coeff) is reported together with the probability (p val) that the coefficient is not relevant for the model. The relative importance of each coefficient (LMG) is computed using the LMG method. }
\label{table11}
\end{table}

\end{document}